\documentclass[preprint,12pt]{elsarticle}

\usepackage{graphicx}

\usepackage{amssymb}
\usepackage{amsmath}
\usepackage{dsfont}
\usepackage{color}

\usepackage[utf8x]{inputenc}

\newcommand{\R}{\mathds{R}}

\newcommand{\dd}{\mathrm{d}}

\usepackage{lineno}

\journal{ }

\begin{document}

\begin{frontmatter}


\title{Localized stationary seismic waves predicted using a nonlinear gradient elasticity model}


\author[label1]{Leo Dostal \corref{cor1}}

\author[label1]{Marten Hollm}

\author[label2]{Andrei V. Metrikine}

\author[label2]{Apostolos Tsouvalas}

\author[label2]{Karel N. van Dalen}

\address[label1]{Institute of Mechanics and Ocean Engineering, Hamburg University of Technology, 21073 Hamburg, Germany}

\address[label2]{Faculty of Civil Engineering \& Geosciences, TU Delft, Delft, Netherlands}


 

\begin{abstract}
This paper aims at investigating the existence of localized stationary waves in the shallow subsurface whose constitutive behaviour is governed by the hyperbolic model, implying non-polynomial nonlinearity and strain-dependent shear modulus. To this end, we derive a novel equation of motion for a nonlinear gradient elasticity model, where the higher-order gradient terms capture the effect of small-scale soil heterogeneity/micro-structure. We also present a novel finite-difference scheme to solve the nonlinear equation of motion in space and time. Simulations of the propagation of arbitrary initial pulses clearly reveal the influence of the nonlinearity: strain-dependent speed in general and, as a result, sharpening of the pulses. Stationary solutions of the equation of motion are obtained by introducing the moving reference frame together with the stationarity assumption. Periodic (with and without a descending trend) as well as localized stationary waves are found by analyzing the obtained ordinary differential equation in the phase portrait, and integrating it along the different trajectories. The localized stationary wave is in fact a kink wave and is obtained by integration along a homoclinic orbit. In general, the closer the trajectory lies to a homoclinic orbit, the sharper the edges of the corresponding periodic stationary wave and the larger its period. Finally, we find that the kink wave is in fact not a true soliton as the original shapes of two colliding kink waves are not recovered after interaction. However, it may have high amplitude and reach the surface depending on the damping mechanisms (which have not been considered). Therefore, seismic site response analyses should not a priori exclude the presence of such localized stationary waves.
\end{abstract}





\end{frontmatter}


\section{Introduction}\label{sec:Introduction}
\label{S:1}
 For the prediction of the so-called seismic site response - the response of the top soil layers of the earth - induced by seismic waves, typically 1-D models are employed \cite{regnier2016international}. The so-called equivalent linear scheme is used very often, in which the soil stiffness and damping are modelled assuming constant shear modulus and material damping ratio, respectively~\cite{rodriguez2017regional}. The actual values of the shear modulus and damping ratio of a specific layer in the soil profile are based on the maximum level of strain observed inside that layer; this requires iteration as it is not possible to determine the maximum strain level a priori. For high maximum strain levels in the soil layers, such equivalent shear modulus and damping ratio cannot accurately represent the behavior over the entire duration of a seismic event, as the strains vary significantly. In such cases, a nonlinear time domain solution is typically used to account for the variation of the shear modulus and damping ratio during shaking (e.g., \cite{regnier2016international}).
 
Contemporary research into nonlinear time-domain models for seismic site response analysis is mostly focused on the development of advanced constitutive models so as to capture important features of the soil behavior such as anisotropy, pore water pressure generation and dilation~\cite{hashash2010recent}. However, limited research has been devoted to the fundamental nonlinear-dynamics aspects of the seismic site response. Employing the commonly used hyperbolic constitutive model \cite{hardin1972shear,kramer1996geotechnical}, softening behaviour and super-harmonic resonances were recently demonstrated for a superficial soil layer under uniform harmonic excitation at the lower boundary~\cite{zhangharmonic}. However, the possibility for localized stationary waves such as solitons to propagate through the soil column and reach the surface has not been widely recognized in the seismological literature and neither in the geo-technical literature, although a number of publications hint at the possibility of Love-type surface solitary waves, solitary waves along faults, and the importance of nonlinearity in general~\cite{2008AGUSMS,bataille1982nonlinear,maugin2007nonlinear,bataille2008seismic,calisto2010evidence,calisto2014envelope,majewski2006seismic,majewski2006tectonic}.

In this paper, we therefore investigate the existence of localized stationary waves in the shallow subsurface with the constitutive behaviour governed by the hyperbolic model, implying that the shear modulus is strain dependent (i.e., non-polynomial nonlinearity). As the classical wave equation with this particular nonlinearity has non-physical discontinuous solutions, we employ a nonlinear \text{\it gradient elasticity} \text{\it model}, which sometimes is also called a higher-order gradient continuum or a micro-structured solid; the equation of motion is of the Boussinesq-type. Compared to the classical continuum, the stress–strain relation has higher-order gradient terms to capture the effect of small-scale soil heterogeneity/ micro-structure (yet keeping the description of the material homogeneous), which introduces dispersive effects particularly for the shorter waves~\cite{metrikine2006causality,mindlin1963microstructure,vardoulakis1994role,rubin1995continuum,muhlhaus1996dispersion,andrianov2011wave,askes2008four,askes2011gradient,papargyri2011transient}. Such higher-order gradient terms are naturally obtained using asymptotic homogenization techniques for periodically inhomogeneous media \cite{metrikine2006causality,fish2002non,andrianov2008higher,craster2010high,capdeville20101}.  Dispersive effects significantly influence the behaviour of localized stationary waves, as such waves exist exactly because of the balance between dispersive and nonlinear effects, allowing their propagation without distortion. The dispersion prohibits the formation of jumps, which leads to physically realisable solutions. So-called higher-order dispersion correction and higher-order dispersive nonlinearity has been discussed in the context of the well-known Korteweg-de Vries equation~\cite{belashov2006solitary}. For higher-order gradient elasticity continua, the existence and properties of localized stationary waves have been studied for polynomial nonlinearity in both the macro-scale and the micro-scale terms (i.e., in the higher-order derivative terms)~\cite{majewski2006seismic,salupere2008numerical,randruut2010one,tamm2012propagation,berezovski2013dispersive,engelbrecht2015reflections,engelbrecht2018solitons,engelbrecht2018nonlinear}. However, such waves have never been studied in the context of higher-order gradient elasticity continua that are dictated by hyperbolic nonlinearity, which is done in this paper.

Sticking to the 1-D assumption, we derive the equation of motion for the nonlinear gradient elasticity model based on Newton’s second law and Eringen's general stress-strain relation~\cite{eringen2002nonlocal}, which allows introducing the higher-order derivatives in an unambiguous manner (Section \ref{sec:modelling}).
An ordinary differential equation from which stationary solutions for the nonlinear gradient elasticity model can be obtained, is derived in Section \ref{sec:StationaryWaveSolutions}. Periodic (with and without a descending trend) as well as localized stationary waves are found (Section \ref{sec:StatWaveSols_results}); the latter is in fact a \textit{kink wave} and is obtained by integrating along a homoclinic orbit in the phase portrait~\cite{rubin1995continuum}, like it can be done for the well-known sine-Gordon equation. \textcolor{black}{Section \ref{sec:Scheme} presents a novel numerical scheme to solve the nonlinear equation of motion in space and time. It exploits the structure of the partial differential equation in order to simplify the computation of the spatial finite-difference approximations. The pseudospectral scheme presented in \cite{salupere2008numerical} cannot be directly applied for the hyperbolic nonlinearity, since the roots of a polynomial with non-integer exponents would have to be determined; this cannot be done analytically and is therefore numerically expensive. For that reason, the introduced numerical scheme directly considers the hyperbolic nonlinearity. Another advantage of the numerical scheme is that, because it employs the finite-difference method, we are not limited to periodic boundary conditions.} The scheme is used to study the propagation of arbitrary initial pulses (Section \ref{sec:ResultsGaussian}) as well as to check the stationarity of the stationary waves identified (Section \ref{sec:StatWaveSols_results}). In addition, it allows to demonstrate that the kink wave is in fact not a true \text{\it soliton}. This is done by means of a numerical collision experiment in which two kink waves propagate in opposite direction and pass each other (Section \ref{sec:KinkInteraction}); after interaction, their original shapes are not recovered. 

Even though not being a true soliton, the kink wave, which may have high amplitude, can propagate through the soil column and potentially reach the surface depending on the strength of the material and geometrical damping mechanisms (which have not been considered). Therefore, seismic site response analyses should not a priori exclude the presence of such localized stationary waves.





\section{Model}\label{sec:modelling}

\subsection{General framework}
The starting point to derive the equation of motion of the gradient elasticity model is Newton's second law. For transverse waves propagating in the vertical direction $z$ and considering the one-dimensional situation, it reads~\cite{verruijt2009introduction} 
\begin{equation}\label{eq:eomstart}
\rho \frac{\partial^2 u}{\partial t^2}=\frac{\partial \sigma_{zx}}{\partial z}.
\end{equation}
%
Here, $\rho$ denotes the material density, $u(z,t)$ the horizontal displacement in the $x$ direction, $t$ is time, and $\sigma_{zx}$ is the shear stress.
In a nonlinear system, the stress-strain relation can generally be written as follows ~\cite{eringen2002nonlocal}:
\begin{equation}\label{eq:stress_strain}
\sigma_{zx}(z,t)=\int_{-\infty}^{\infty}\int_{-\infty}^{\infty} g\left(z-\zeta, t-\tau, \gamma(\zeta,\tau)\right)\varepsilon_{zx}(\zeta,\tau) \dd \zeta \dd \tau.
\end{equation}
This relation expresses that the stress $\sigma_{zx}$ at location $z$ and time $t$ generally depends on the strain $\varepsilon_{zx}$ at all points of the. Moreover, on the entire strain history (note that $\zeta$ and $\tau$ denote auxiliary space and time variables, respectively); the specific nonlocality and history dependence is contained in the kernel function $g(z,t,\gamma)$. The strain depends on the displacement in the following way~\cite{verruijt2009introduction}:
\begin{equation}\label{eq:epsilonzx}
\varepsilon_{zx}=\frac{1}{2}\frac{\partial u}{\partial z}.
\end{equation}
The nonlinearity comes into play through the $\gamma$ dependence of the kernel function, which expresses that the elasticity operators, i.e. the elastic coefficients, are strain dependent. In this work, we relate $\gamma$ to the deviatoric component of the so-called octahedral strain, which for the one-dimensional case leads to~\cite{polakowski1966strength,Zhang}
\begin{equation}\label{eq:gamma}
 \gamma=\sqrt{3}|\varepsilon_{zx}|=\frac{\sqrt{3}}{2} \left|\frac{\partial u}{\partial z}\right|.
\end{equation}



\subsection{Equation of motion for nonlinear gradient elasticity model}
\label{subsec:nonlinearMedium}
If the distribution of the kernel function $g(z,t,\gamma)$ in $z$ and $t$ is arbitrary, the equation of motion, as obtained by substituting Eq.~\ref{eq:stress_strain}) into  Eq.~(\ref{eq:eomstart}), will be of the integro-differential type. However, as our aim is to derive the equation of motion for a gradient elasticity model, which is a partial differential equation, we restrict ourselves to Dirac delta functions. Apart from the conventional term, we include terms with double space and double time derivatives: 
%
%
\begin{equation}\label{eq:kernelgNonlin}
\begin{aligned}
g(z-\zeta, t-\tau, \gamma)= &\, 2\big(G(\gamma)\delta(z-\zeta)\delta(t-\tau)-L^2 G^{(L)}(\gamma)\delta_{,\zeta\zeta}(z-\zeta)\delta(t-\tau)\\
&  +T^2 G^{(T)}(\gamma)\delta(z-\zeta)\delta_{,\tau\tau}(t-\tau) \big),    
\end{aligned}
\end{equation}
where $\delta(...)$ denotes Dirac's delta function, and $(...)_{,\zeta\zeta}$ and $(...)_{,\tau\tau}$ denote double partial differentiation with respect to $\zeta$ and $\tau$, respectively (this notation is used throughout the paper, when useful, to denote partial derivatives); $G(\gamma)$ is the strain-dependent shear modulus, $G^{(L)}(\gamma)$ and $G^{(T)}(\gamma)$ are additional strain-dependent elastic moduli (related to higher-order derivative terms, as shown below), and $\gamma=\gamma(\zeta,\tau)$. The semi-local operators with derivatives of the Dirac function come with the time $(T)$ and length $(L)$ scales that characterize the specific history dependence and nonlocality of the medium, respectively; it is well-known that particular nonlocal and history effects can indeed be captured by using higher-order derivative terms in the equation of motion (e.g., \cite{benvenuti2013one}). The kernel function with the two additional terms (with double space and double time derivatives) and corresponding signs is chosen in accordance with the linear model (see Eq. \eqref{eq:linEoM1} below), for which it ensures unconditional stability as well as realistic lower and upper bounds for the speed of energy transfer of the propagating wave \cite{metrikine2006causality}.

Now, inserting Eq.~\eqref{eq:kernelgNonlin} into the Eq.~\eqref{eq:stress_strain}, we obtain the stress-strain relation of the nonlinear gradient elasticity model:
%
\textcolor{black}{
\begin{equation}\label{eq:stress_strainHO}
\begin{aligned}
\sigma_{zx}(z,t)=& 2\,\biggl( G(\gamma)\varepsilon_{zx}(z,t)-L^2 \frac{\partial^2}{\partial z^2}\left( G^{(L)}(\gamma)\varepsilon_{zx}(z,t)\right)\\ 
&+T^2 \frac{\partial^2}{\partial t^2}\left( G^{(T)}(\gamma)\varepsilon_{zx}(z,t)\right) \biggl ).
\end{aligned}
\end{equation}
}%
Substituting the Eqs. \eqref{eq:epsilonzx} and \eqref{eq:stress_strainHO} into Eq. \eqref{eq:eomstart} yields the corresponding equation of motion: 
\begin{equation}\label{eq:nonlinEoM1}
\rho \frac{\partial^2 u}{\partial t^2}=\frac{\partial}{\partial z} \left(G(\gamma) \frac{\partial u}{\partial z}-L^2 \frac{\partial^2}{\partial z^2}\left( G^{(L)}(\gamma)\frac{\partial u}{\partial z} \right) +T^2 \frac{\partial^2}{\partial t^2}\left( G^{(T)}(\gamma)\frac{\partial u}{\partial z}\right) \right).
\end{equation}

For simplicity, we now relate the strain-dependent additional elastic moduli to the conventional strain-dependent shear modulus $G(\gamma)$ using dimensionless constants $B_1$ and $B_2$:
\begin{equation}\label{eq:elasticParametersG}
\begin{aligned}
    G^{(L)}(\gamma)&=B_1 G(\gamma),\\
    G^{(T)}(\gamma)&=B_2 G(\gamma).
\end{aligned}
\end{equation}
Without loss of generality, we interrelate the characteristic length and time scales:
\begin{equation}\label{eq:relation_TL}
    T^2 = \frac{L^2}{c_0^2}.
\end{equation}
Here, $c_0^2=G_0/\rho$, with $G_0$ being the well-known small-strain shear modulus from linear elasticity; $c_0$ is the corresponding shear-wave speed. Tthroughout the paper, the subscript “$0$” indicates that the quantity relates to small-strain/linear behavior. Using Eqs. \eqref{eq:elasticParametersG} and \eqref{eq:relation_TL}, Eq.~\eqref{eq:nonlinEoM1} can be written as 
\begin{equation}\label{eq:nonlinEoM2}
\rho \frac{\partial^2 u}{\partial t^2} =\frac{\partial}{\partial z} \left(G(\gamma) \frac{\partial u}{\partial z}-B_1 L^2 \frac{\partial^2}{\partial z^2}\left( G(\gamma)\frac{\partial u}{\partial z}\right) +B_2\frac{\rho L^2}{G_0} \frac{\partial^2}{\partial t^2}\left( G(\gamma)\frac{\partial u}{\partial z} \right) \right),
\end{equation}
%
which is the final form of the equation of motion of the nonlinear gradient elasticity model.

In this work, we use the hyperbolic soil model typically employed for seismic site response analyses, with the following expression for the strain-dependent shear modulus \cite{hardin1972shear}:   
\begin{equation}\label{eq:haperbolicSoil}
    G(\gamma)=\frac{G_0}{1+\left(\gamma / \gamma_{\mathrm{ref}}  \right)^{\beta}}.
\end{equation}
Here, $\gamma_{\mathrm{ref}}$ denotes a reference shear strain, and $\beta$ is a dimensionless constant ($0<\beta<1)$.

\subsection{Limit case}

For completeness, we here consider the limit case of $\gamma/\gamma_{\mathrm{ref}} \rightarrow 1$, which reduces the model to a linear gradient elasticity model. In that case (cf. Eq. \eqref{eq:elasticParametersG})
\begin{equation}\label{eq:elasticParametersG_0}
\begin{aligned}
    G^{(L)}&=G_0^{(L)}=B_1 G_0,\\
    G^{(T)}&=G_0^{(T)}=B_2 G_0.
\end{aligned}
\end{equation}
Using the same definition for $T$ (Eq.~\eqref{eq:relation_TL}), the stress-strain relation~Eq.~\eqref{eq:stress_strainHO} reduces to
\begin{equation}\label{eq:stress_strainHOlinear}
\sigma_{zx}=2\left(G_0\varepsilon_{zx} -  B_1 G_0 L^2 \frac{\partial^2\varepsilon_{zx}}{\partial z^2} + B_2 \rho L^2 \frac{\partial^2\varepsilon_{zx}}{\partial t^2} \right),
\end{equation}
whereby .
The corresponding equation of motion reads as follows:
\begin{equation}\label{eq:linEoM1}
\rho \frac{\partial^2 u}{\partial t^2}=G_0\frac{\partial^2 u}{\partial z^2} - B_1 G_0 L^2 \frac{\partial^4 u}{\partial z^4}  + B_2 \rho L^2 \frac{\partial^4 u}{\partial z^2\partial t^2}.
\end{equation}
This linear equation was derived before by Metrikine and Askes \cite{metrikine2002one} from a discrete model using a continualization procedure. A similar equation was used by Georgiadis~et~al.~\cite{georgiadis2000torsional} to investigate the existence of horizontally polarized surface waves.

\section{Stationary wave solutions}\label{sec:StationaryWaveSolutions}
It is possible to determine stationary solutions of the equation of motion Eq.~\eqref{eq:nonlinEoM2} that do not change their shape while propagating in the nonlinear medium. This means that, starting at an initial condition, the stationary solution does not change with respect to the coordinate $\xi=z-ct$, which moves with the velocity $c\in \R$. Thereby, $c$ can be arbitrarily chosen and leads to a specific solution. In order to find possible stationary solutions, we apply  
the transformation $\xi=z-ct$ and assume stationarity, which yields
\begin{equation}\label{eq:xiTransform}
    u_{,tt}=u_{,\xi\xi}c^2,\quad u_{,ztt}=u_{,\xi\xi\xi}c^2,\quad u_{,zt}=-u_{,\xi\xi}c,\quad \frac{\partial }{\partial z}=\frac{\partial }{\partial \xi}.
\end{equation}
Here, we recall that these particular subscripts (time or space variable(s) preceded by a comma) denote partial derivative operators. Substituting expressions of Eq.~\eqref{eq:xiTransform} into Eq.~\eqref{eq:nonlinEoM2}, we get an ordinary differential equation for the determination of stationary solutions to Eq.~\eqref{eq:nonlinEoM2}: 
\begin{equation}\label{eq:nonlinEoMSoliton}
\rho\,c^2\, u_{,\xi\xi}=\frac{\partial}{\partial \xi} \left(G(\gamma) u_{,\xi} +\left(B_2\,c^2\,\frac{\rho L^2}{G_0}- B_1 L^2 \right) \frac{\partial^2}{\partial \xi^2}\left( G(\gamma)u_{,\xi}\right) \right).
\end{equation}
%
\textcolor{black}{
In order to evaluate $\frac{\partial^2}{\partial \xi^2}\left( G(\gamma)u_{,\xi}\right)$ in Eq.~\eqref{eq:nonlinEoMSoliton} for the hyperbolic soil model from Eq.~\eqref{eq:haperbolicSoil}, it has to be noted that the absolute value function $g(x)=\vert x \vert$ is not differentiable for $x=0$. However, it is weakly differentiable with sgn($x$), which denotes the sign function, as weak derivative. Using this, we specifically obtain}

\begin{equation}\label{eq:SolitonSpecific}
\begin{aligned}
    \rho c^2 u_{,\xi\xi}=&\frac{\partial }{\partial \xi}\Biggl[ u_{,\xi}\frac{G_0}{1+\left(\frac{\sqrt{3}|u_{,\xi}| }{2\gamma_{\mathrm{ref}}}  \right)^{\beta}}+\left(B_2\,c^2\,\rho L^2-G_0 B_1 L^2 \right)\Biggl\{
     u_{,\xi\xi\xi}\Bigg[\frac{1}{1+\left(\frac{\sqrt{3}|u_{,\xi}| }{2\gamma_{\mathrm{ref}}}  \right)^{\beta}}\\
    &-\beta \left(\frac{\sqrt{3}|u_{,\xi}| }{2\gamma_{\mathrm{ref}}}  \right)^{\beta} \left(1+\left(\frac{\sqrt{3}|u_{,\xi}| }{2\gamma_{\mathrm{ref}}}  \right)^{\beta}\right)^{-2}
    \Bigg] \\
    &+\frac{\sqrt{3}\,\mathrm{sgn}(u_{,\xi})}{2\gamma_{\mathrm{ref}}}u_{,\xi\xi}^2\Bigg(2\beta^2 \left(\frac{\sqrt{3}|u_{,\xi}| }{2\gamma_{\mathrm{ref}}}  \right)^{2\beta-1}\left(1+\left(\frac{\sqrt{3}|u_{,\xi}| }{2\gamma_{\mathrm{ref}}}  \right)^{\beta}\right)^{-3}\\
    &-(\beta+\beta^2) \left(\frac{\sqrt{3}|u_{,\xi}| }{2\gamma_{\mathrm{ref}}}  \right)^{\beta-1}\left(1+\left(\frac{\sqrt{3}|u_{,\xi}| }{2\gamma_{\mathrm{ref}}}  \right)^{\beta}\right)^{-2}
    \Bigg)
    \Biggl\}
    \Biggl].
    \end{aligned}
\end{equation}
Integration of Eq.~\eqref{eq:SolitonSpecific} with respect to $\xi$, grouping terms related to $u_{,\xi\xi\xi}$, and setting $y:=u_{,\xi}$, $\Dot{()}:=\frac{\partial }{\partial \xi}$
yields
\begin{equation}\label{eq:SolitonSpecificIntegrated}
\begin{aligned}
\Ddot{y}=&\frac{1}{1-\beta \left(\frac{\sqrt{3}|y| }{2\gamma_{\mathrm{ref}}}  \right)^{\beta} \left(1+\left(\frac{\sqrt{3}|y| }{2\gamma_{\mathrm{ref}}}  \right)^{\beta}\right)^{-1}} \Biggl\{\frac{\rho c^2\left(1+\left(\frac{\sqrt{3}|y| }{2\gamma_{\mathrm{ref}}}  \right)^{\beta}\right)-G_0}{B_2 c^2\rho L^2-G_0B_1L^2 }y\\
&-\frac{\sqrt{3}\,\mathrm{sgn}(y)}{2\gamma_{\mathrm{ref}}} \Biggl[
2\beta^2 \left(\frac{\sqrt{3}|y| }{2\gamma_{\mathrm{ref}}}  \right)^{2\beta-1}\left(1+\left(\frac{\sqrt{3}|y| }{2\gamma_{\mathrm{ref}}}  \right)^{\beta}\right)^{-2}\\
&-(\beta+\beta^2) \left(\frac{\sqrt{3}|y| }{2\gamma_{\mathrm{ref}}}  \right)^{\beta-1}\left(1+\left(\frac{\sqrt{3}|y| }{2\gamma_{\mathrm{ref}}}  \right)^{\beta}\right)^{-1}
\Biggl] \Dot{y}^2
\Biggl\}.
    \end{aligned}
\end{equation}
This is a nonlinear second-order ordinary differential equation, which can be solved for stationary wave solutions of Eq.~\eqref{eq:nonlinEoM2}. Specific stationary solutions are determined in Section~\ref{sec:results}.

\section{Numerical scheme}\label{sec:Scheme}
In order to solve Eq.~\eqref{eq:nonlinEoM2} in space and time,
we use a newly developed fully implicit scheme for the numerical solution of partial differential equations of the form
\begin{equation}\label{eq:GeneralNonlinEoM}
    \begin{aligned}
        \rho\frac{\partial^2 u }{\partial t^2}=&\frac{\partial}{\partial z} \biggl(G\left( \frac{\partial u}{\partial z}\right) \frac{\partial u}{\partial z}-B_1 L^2 \frac{\partial^2}{\partial z^2}\left( G\left(\frac{\partial u}{\partial z}\right)\frac{\partial u}{\partial z}\right)\\ &
        +B_2\frac{\rho L^2}{G_0} \frac{\partial^2}{\partial t^2}\left( G\left(\frac{\partial u}{\partial z}\right)\frac{\partial u}{\partial z}\right) \biggl),
    \end{aligned}
\end{equation}
where $G\left( \frac{\partial u}{\partial z}\right)$ denotes that $G$ can depend on $\partial u/\partial z$ in an arbitrary manner.
%
It is assumed that the solution $u(z,t)$ of Eq. (\ref{eq:nonlinEoM2}) exists in time $t \in [0,\mathbb{T}]$ and space $z \in [Z_{\ell},Z_{h}]$. Therefore, a grid in time
\begin{equation}
    0=t_0<t_1<\dots<t_N=\mathbb{T}, \quad t_n=n \Delta t \; \textup{ for }n=0,\dots, N, \; \Delta t=\frac{\mathbb{T}}{N}, 
\end{equation}
and a grid in space
\begin{equation}
    Z_{\ell}=z_0<z_1<\dots<z_M=Z_{h}, \quad z_i=i \Delta z \; \textup{ for }i=0,\dots, M, \; \Delta z=\frac{Z_{\ell}-Z_{h}}{M}, 
\end{equation}
are introduced. Using
\begin{equation} \label{eq:nonlin_h}
    h(u_{,z}):=G(\textcolor{black}{u_{,z}}) \, u_{,z},
\end{equation}
Eq. (\ref{eq:GeneralNonlinEoM}) can be written as
\begin{equation} \label{eq:eq_in_h}
    \rho \, u_{,tt}=h_{,z} -B_1 L^2 h_{,zzz} + B_2 \frac{\rho L^2}{G_0} h_{,ttz}.
\end{equation}
\textcolor{black}{By using Eq. (\ref{eq:nonlin_h}) and (\ref{eq:eq_in_h})}, the structure of the considered partial differential equation is exploited. \textcolor{black}{This will simplify the computation} of the spatial finite difference approximations\textcolor{black}{, as can be seen in the further course of this section}. Assuming that the solution is known at the timepoints $t_{n-1}$ and $t_{n}$ and replacing the time derivative by a finite-difference approximation, Eq.~(\ref{eq:eq_in_h}) results in the nonlinear equation
\begin{equation} \label{eq:nonlin_f}
    \mathbf{f}(\mathbf{u}^{n+1})=\mathbf{0},
\end{equation}
with
\begin{equation}
    \mathbf{u}^{n+1}:=\left[u_0^{n+1}, u_1^{n+1}, \dots, u_M^{n+1} \right]^{\mathrm{T}},
\end{equation}
and
\begin{equation} \label{eq:nonlin_f_def}
\begin{aligned}
f_i(\mathbf{u}^{n+1}):=&\rho \, \frac{u_i^{n+1}-2u_i^{n}+u_i^{n-1}}{\Delta t^2}-\textcolor{black}{\frac{h_{,z}(u_{i,z}^{n+1})+2h_{,z}(u_{i,z}^{n})+h_{,z}(u_{i,z}^{n-1})}{4}} \\
&+B_1 L^2\textcolor{black}{\frac{h_{,zzz}(u_{i,z}^{n+1})+2h_{,zzz}(u_{i,z}^{n})+h_{,zzz}(u_{i,z}^{n-1})}{4}}\\
&- B_2 \frac{\rho L^2}{G_0} \frac{h_{,z}(u_{i,z}^{n+1})-2h_{,z}(u_{i,z}^{n})+h_{,z}(u_{i,z}^{n-1})}{\Delta t^2},
    \end{aligned}
\end{equation}
where $f_i$ is the $i$-th component of $\mathbf{f}$, $u_i^{n}$ is a grid function approximating the solution at time $t_n$ and space $z_i$, i.e. $u_i^{n} \approx u(z_i,t_n)$, and $u_{i,z}^{n}$ approximates $u_{,z}(z_i,t_n)$. \textcolor{black}{Equations (\ref{eq:nonlin_f}) and (\ref{eq:nonlin_f_def}) approximate Eq. (\ref{eq:eq_in_h}) (evaluated at $t=t_n$) up to an accuracy of $\mathcal{O}(\Delta t^2)$, whereby $\mathcal{O}(\cdot)$ represents the big O notation.}
Next, the space derivatives are discretized. For this, standard finite-difference approximations are used again. In order to simplify the notation, the time index $n$ is omitted in the following, such that $\mathbf{u}^n$ is written as $\mathbf{u}$ or $u_i^{n}$ is written as $u_i$. This leads to the approximations\textcolor{black}{, which have all an accuracy of $\mathcal{O}(\Delta z^2)$}:
\begin{equation} \label{eq:approx_uz}
    u_{i,z}=\frac{u_{i+1}-u_{i-1}}{2\Delta z} \textcolor{black}{+\mathcal{O}(\Delta z^2)},
\end{equation}
\begin{equation} \label{eq:approx_hz}
    \begin{aligned}
        h_{,z}(u_{i,z})=&\frac{h(u_{i+1,z})-h(u_{i-1,z})}{2\Delta z}\textcolor{black}{+\mathcal{O}(\Delta z^2)}\\
        =& \frac{1}{2 \Delta z} \biggl \lbrace \frac{G_0}{1+\left( \frac{\sqrt{3}}{2} \frac{\vert u_{i+2}-u_i \vert}{2 \Delta z \gamma_{\mathrm{ref}}} \right)^\beta} \frac{u_{i+2}-u_i}{2 \Delta z}-\frac{G_0}{1+\left( \frac{\sqrt{3}}{2} \frac{\vert u_{i}-u_{i-2} \vert}{2 \Delta z \gamma_{\mathrm{ref}}} \right)^\beta} \frac{u_{i}-u_{i-2}}{2 \Delta z} \biggl \rbrace\\
        &\textcolor{black}{+\mathcal{O}(\Delta z^2)},
    \end{aligned}
\end{equation}
\begin{equation} \label{eq:approx_hzzz}
    \begin{aligned}
        h_{,zzz}(u_{i,z})=&\frac{h(u_{i+2,z})-2h(u_{i+1,z})+2h(u_{i-1,z})-h(u_{i-2,z})}{2\Delta z^3}\textcolor{black}{+\mathcal{O}(\Delta z^2)}\\
        =& \frac{1}{2 \Delta z^3}\biggl \lbrace \frac{G_0}{1+\left( \frac{\sqrt{3}}{2} \frac{\vert u_{i+3}-u_{i+1} \vert}{2 \Delta z \gamma_{\mathrm{ref}}} \right)^\beta} \frac{u_{i+3}-u_{i+1}}{2 \Delta z} \\
        &-2\frac{G_0}{1+\left( \frac{\sqrt{3}}{2} \frac{\vert u_{i+2}-u_i \vert}{2 \Delta z \gamma_{\mathrm{ref}}} \right)^\beta} \frac{u_{i+2}-u_i}{2 \Delta z}+2\frac{G_0}{1+\left( \frac{\sqrt{3}}{2} \frac{\vert u_{i}-u_{i-2} \vert}{2 \Delta z \gamma_{\mathrm{ref}}} \right)^\beta} \frac{u_{i}-u_{i-2}}{2 \Delta z} \\
        &-\frac{G_0}{1+\left( \frac{\sqrt{3}}{2} \frac{\vert u_{i-1}-u_{i-3} \vert}{2 \Delta z \gamma_{\mathrm{ref}}} \right)^\beta} \frac{u_{i-1}-u_{i-3}}{2 \Delta z} \biggl \rbrace\textcolor{black}{+\mathcal{O}(\Delta z^2)}.
    \end{aligned}
\end{equation}

\textcolor{black}{Now, the advantage of the exploitation of the structure of Eq. (\ref{eq:GeneralNonlinEoM}) and the introduction of the function $h(u_{,z})$ can be seen. In Eq. (\ref{eq:GeneralNonlinEoM}) the third-order derivative of $G(u_{,z})$ with respect to $z$ appears. However if the hyperbolic soil model $G(\gamma)$ from Eq.~\eqref{eq:haperbolicSoil} is used for $G(u_{,z})$ in Eq.~\eqref{eq:GeneralNonlinEoM}, then $G(u_{,z})=G(\gamma)$ contains the absolute value function, which is only one time weakly differentiable. A direct \textcolor{black}{finite difference approximation of} the third-order derivative of $G(u_{,z})$ must be able to deal with this problem. By using Eq. (\ref{eq:approx_hz}) and (\ref{eq:approx_hzzz}), the missing differentiability in the case $G(u_{,z})=G(\gamma)$ is circumvented.}

Since Eq. (\ref{eq:nonlin_f}) is nonlinear, a numerical scheme has to be used in order to calculate $u_i^{n+1}$ iteratively. For this, Newton's method is used, for which the computation of the Jacobian matrix is necessary. 

\textcolor{black}{Using $|x|_{,x}=$ sgn$(x)$, sgn$(x)x=\vert x \vert$ and defining
\begin{equation}
    k(x,y):=\frac{1+(1-\beta) \left( \dfrac{\sqrt{3}}{2} \dfrac{\vert x-y \vert}{2 \Delta z \gamma_{\mathrm{ref}}} \right)^\beta}{\textcolor{black}{16} \Delta z^4\left(1+\left( \dfrac{\sqrt{3}}{2} \dfrac{\vert x-y \vert}{2 \Delta z \gamma_{\mathrm{ref}}} \right)^\beta \right) ^2 },
\end{equation}
it follows that
\begin{equation}
    \begin{aligned}
        \frac{\partial f_i}{\partial u_{i+3}}\,(\mathbf{u})=& B_1 L^2 G_0 k(u_{i+3},u_{i+1}),\\
        \dfrac{\partial f_i}{\partial u_{i-3}}\,(\mathbf{u})=& B_1 L^2 G_0 k(u_{i-1},u_{i-3}),\\
        \frac{\partial f_i}{\partial u_{i+2}}\,(\mathbf{u})=& -G_0\left(\Delta z^2+\textcolor{black}{4}B_2 \dfrac{\rho L^2 \Delta z^2}{G_0 \Delta t^2} +2 B_1 L^2\right) k(u_{i+2},u_{i}),\\
        \frac{\partial f_i}{\partial u_{i-2}}\,(\mathbf{u})=& -G_0\left(\Delta z^2+\textcolor{black}{4}B_2 \dfrac{\rho L^2 \Delta z^2}{G_0 \Delta t^2} +2 B_1 L^2 \right) k(u_{i},u_{i-2}),\\
        \frac{\partial f_i}{\partial u_{i}}\,(\mathbf{u})=&\frac{\rho}{\Delta t^2}-\frac{\partial f_i}{\partial u_{i+2}}\,(\mathbf{u})-\frac{\partial f_i}{\partial u_{i-2}}\,(\mathbf{u}),\\
        \frac{\partial f_i}{\partial u_{i+1}}\,(\mathbf{u})=&-\frac{\partial f_i}{\partial u_{i+3}}\,(\mathbf{u}),\\
        \frac{\partial f_i}{\partial u_{i-1}}\,(\mathbf{u})=&-\frac{\partial f_i}{\partial u_{i-3}}\,(\mathbf{u}),\\
        \frac{\partial f_i}{\partial u_{j}}\,(\mathbf{u})=&\,\,0 \quad \quad \textup{for } |i-j|>3.
    \end{aligned}
\end{equation}}

\textcolor{black}{This concludes our new numerical scheme for the computation of solutions to Eq.~\eqref{eq:GeneralNonlinEoM}. In order to simulate the solution on an open domain, absorbing boundary conditions are used. The numerical solution of the linear equation of motion (Eq.~\eqref{eq:linEoM1}) can be computed in a similar manner (using a finite difference scheme). For details, see appendix~\ref{Appendix:SchemeSeismicWaves_Linear}.}

\section{Numerical results}\label{sec:results}
In this section we show and discuss numerical results for the nonlinear Eq.~\eqref{eq:nonlinEoM2} with the choice of parameter values indicated in Tab.~\ref{tab:mediumParameters}. The values of $G_0$, $\rho$, $\beta$ and $\gamma_{\text{ref}}$ have been chosen to represent soil, and the values of the parameters related to the higher-order gradient terms are similar to the ones used in~\cite{metrikine2006causality}. 
\textcolor{black}{
\begin{table}[h]
    \centering
    \begin{tabular}{c|c|c|c|c|c|c}
        $G_0$ $\left[ \textup{Pa} \right]$& $\rho$ $\left[ \textup{kg}\,\textup{m}^{-3} \right]$ & $\beta$ $\left[ \textup{-} \right] $ & $\gamma_{\mathrm{ref}}$ $\left[ \textup{-} \right] $& $B_1$ $\left[ \textup{-} \right] $ & $B_2$ $\left[ \textup{-} \right] $& $L$ $\left[ \textup{m} \right]$ \\
        \hline
        $ 111.86\, \cdot \, 10^6$ & $2009.8$ & $0.91$ & $10^{-3}$ & $1$ & $1.78$ & $0.2$
    \end{tabular}
    \caption{Medium parameter values.}
    \label{tab:mediumParameters}
\end{table}
}

 

%
\subsection{Solutions for a Gaussian pulse} \label{sec:ResultsGaussian}
\begin{figure}[t]
	\centering
		\includegraphics[scale = 0.27]{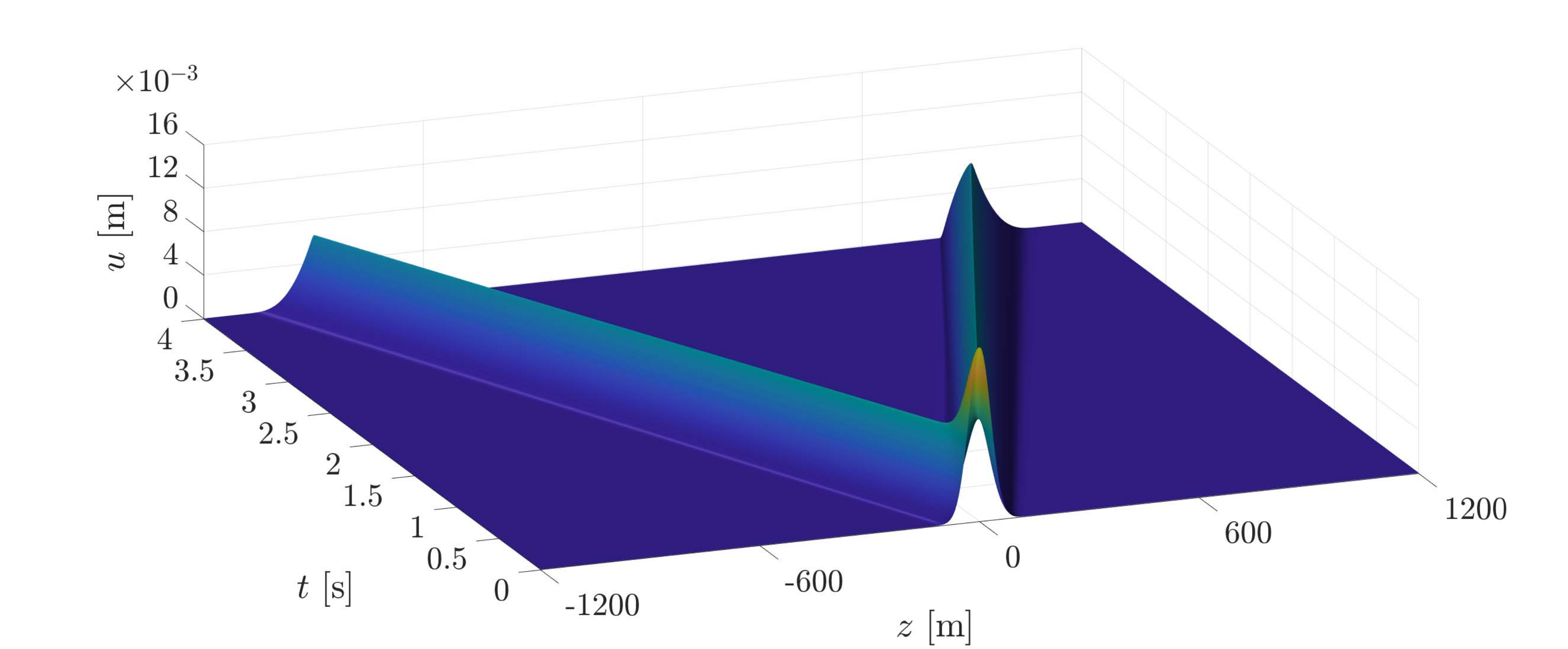}
    \caption{Solution of the nonlinear Eq.~\eqref{eq:nonlinEoM2} with a Gaussian pulse as initial condition with an amplitude of $u_0=0.016\,\textup{m}$.}
	\label{fig:GaussianImpulse}
\end{figure}

First of all, the temporal evolution of a specific solution is studied, where as initial condition a Gaussian pulse is used, i.e. 
\begin{equation}
    u(z,t=0)=u_0 \exp\left(-\frac{z^2}{2\sigma^2}\right).
\end{equation}
Here, the amplitude $u_0=0.016\,\textup{m}$ and standard deviation $\sigma=30\, \textup{m}$ have been chosen to obtain a relatively high strain level, in accordance with \cite{regnier2016international}. In order to compute the numerical solution using the scheme described in Section\,\ref{sec:Scheme}, an initial condition at time point $t_{-1}$ (i.e. $u^{-1}$) has to be chosen. In this study, $u^{-1}=u^{0}$ is used, resulting in a solution with zero initial velocity. The resulting numerical solution can be seen in Fig.~\ref{fig:GaussianImpulse}. The initial pulse divides into two parts, which propagate in opposite directions. We can also observe that the nonlinear solution does not have sharp edges, which would be the case if the higher-order derivative terms were omitted in Eq.~\eqref{eq:nonlinEoM2}; hence the gradient elasticity model yields physically admissible behaviour. 
%

\begin{figure}[t]
	\centering
		\includegraphics[scale = 0.27]{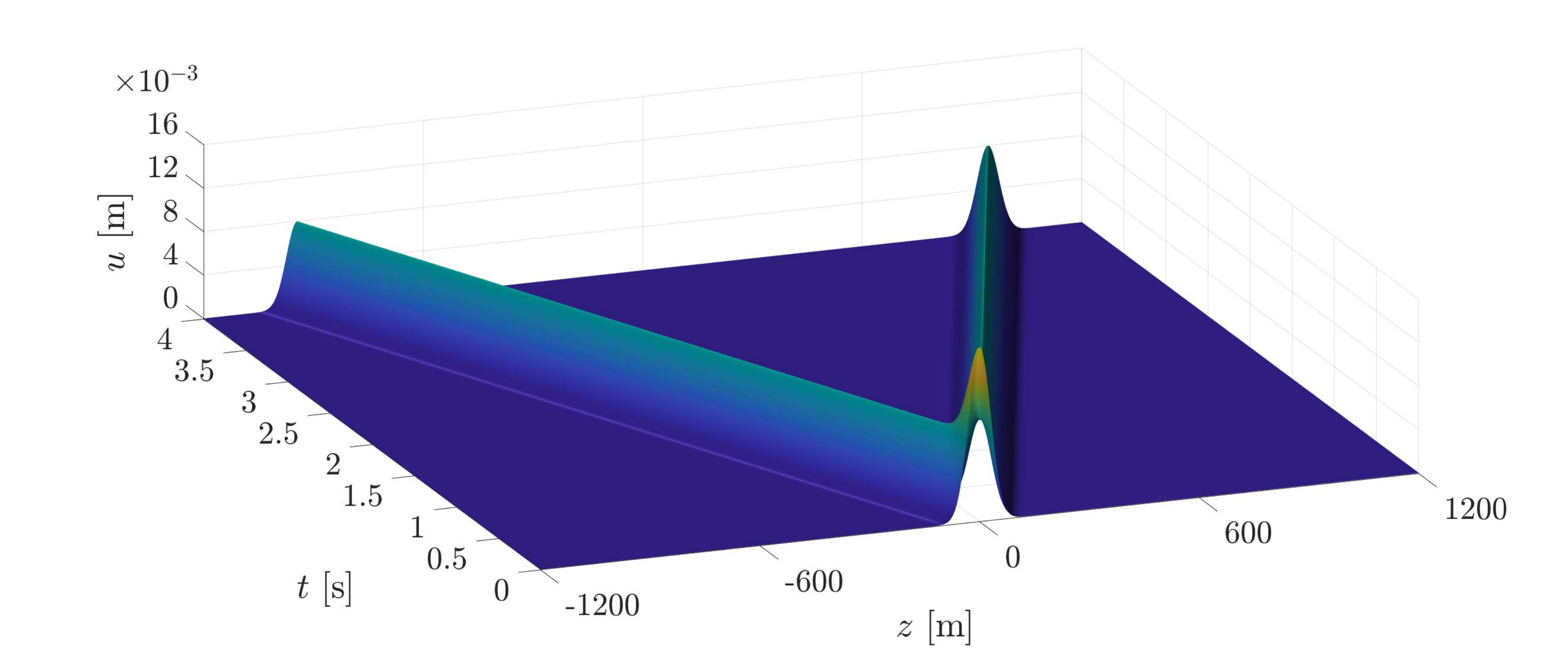}
    \caption{Solution of the linear Eq.~\eqref{eq:linEoM1} with a Gaussian pulse as initial condition with an amplitude of $u_0=0.016\,\textup{m}$.}
	\label{fig:GaussianImpulseLinear}
\end{figure}

\begin{figure}[t]
	\centering	
	\includegraphics[scale = 0.36]{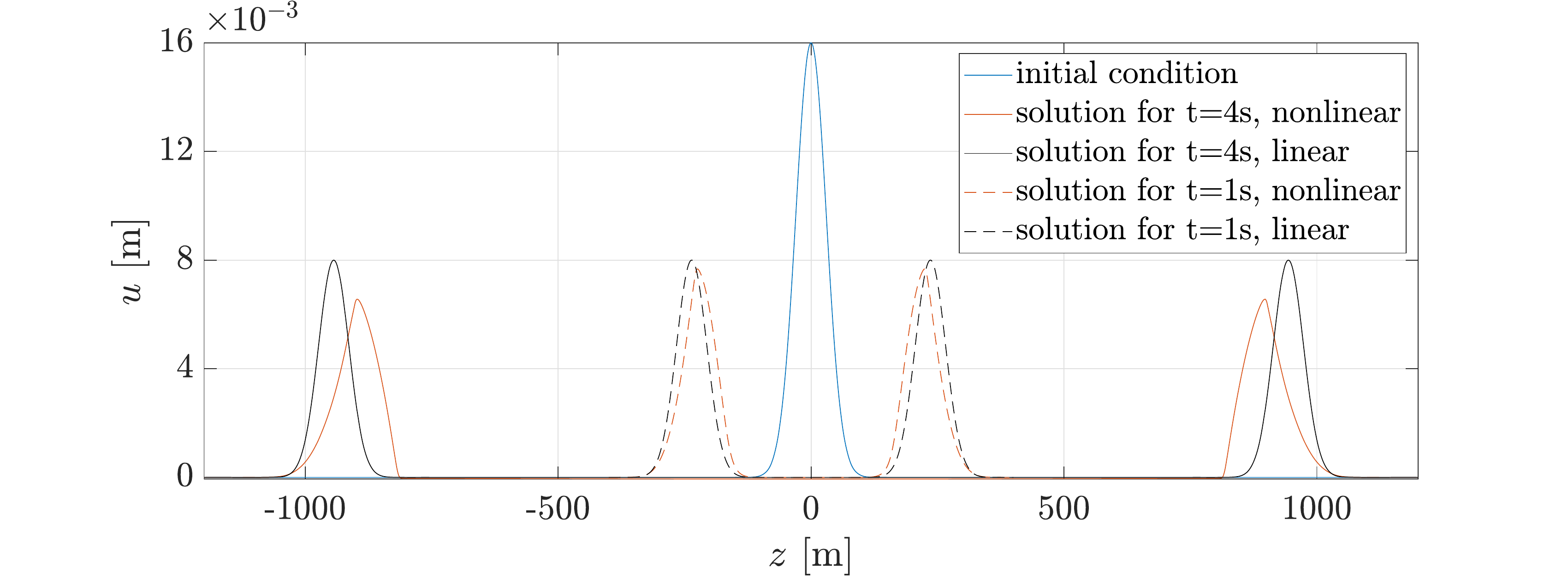}%
    \caption{Comparison of the solution of the linear Eq.~\eqref{eq:linEoM1} and nonlinear Eq.~\eqref{eq:nonlinEoM2} at the end of the simulation time. For both cases, the same initial condition has been used.}
	\label{fig:compLinNonlin}%
\end{figure}

In order to see the effect of the nonlinear terms in Eq.\,\eqref{eq:nonlinEoM2}, the temporal evolution of the corresponding solution for the linear case (Eq.~(\ref{eq:linEoM1})) is shown in Fig.~\ref{fig:GaussianImpulseLinear}. 
There is a clear difference between the linear and nonlinear solutions.
It can be seen that, although the general behavior of the solution stays the same, the nonlinear solution is much sharper compared to the linear solution (but still smooth). In order to make this more clear, Fig.~\ref{fig:compLinNonlin} shows both solutions at the end of the simulation time. 

\begin{figure}[t]
	\centering
	\includegraphics[scale = 0.27]{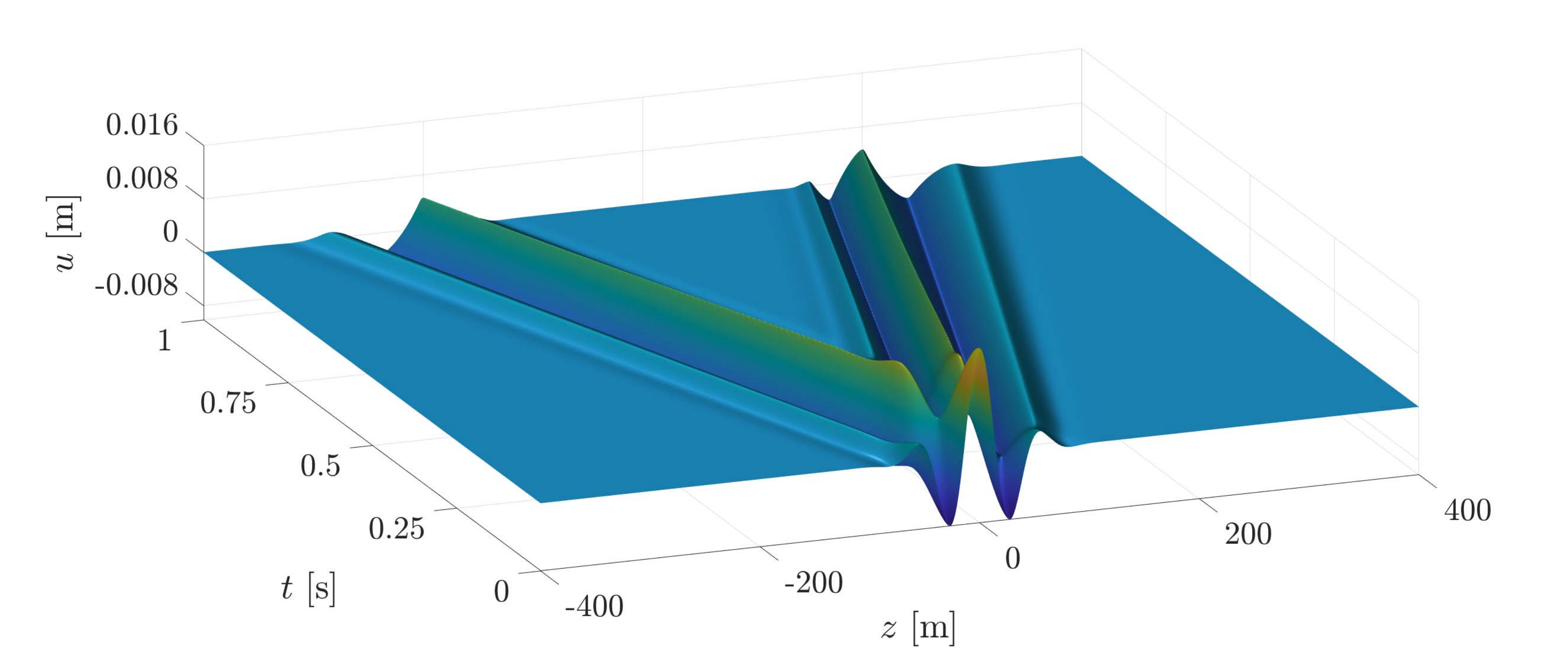}
	\includegraphics[scale = 0.36]{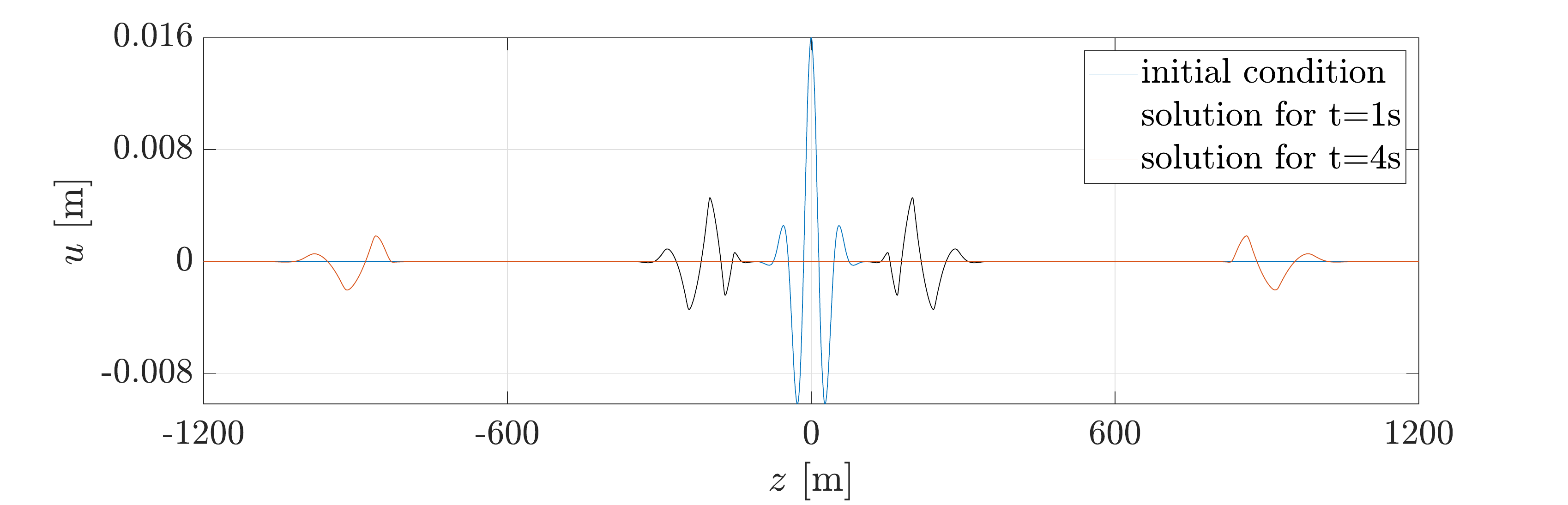}	\caption{Solution of the nonlinear Eq.~\eqref{eq:nonlinEoM2} with a Gaussian-cosine pulse as initial condition with an amplitude of $u_0=0.016\,\textup{m}$.}
	\label{fig:CosNonlin}
\end{figure}

\textcolor{black}{Another effect of the nonlinearity can be observed in Fig. \ref{fig:CosNonlin}, which shows the solution of the nonlinear equation of motion Eq.~\eqref{eq:nonlinEoM2} with a Gaussian-cosine pulse as initial condition, which is given by
\begin{equation}
    u(z,t=0)=u_0 \exp\left(-\frac{z^2}{2\sigma^2}\right)\cos\left(\frac{\pi}{\sigma}z\right).
\end{equation}
Thereby, an amplitude of $u_0=0.016\,\textup{m}$ and standard deviation of $\sigma=30\,\textup{m}$ have been chosen.
It can be well observed that, due to the nonlinearity, different parts of the pulses having different slopes (i.e., strain levels), travel with different speed.}

\begin{figure}[h!]
	\centering	
 	\includegraphics[scale = 0.27]{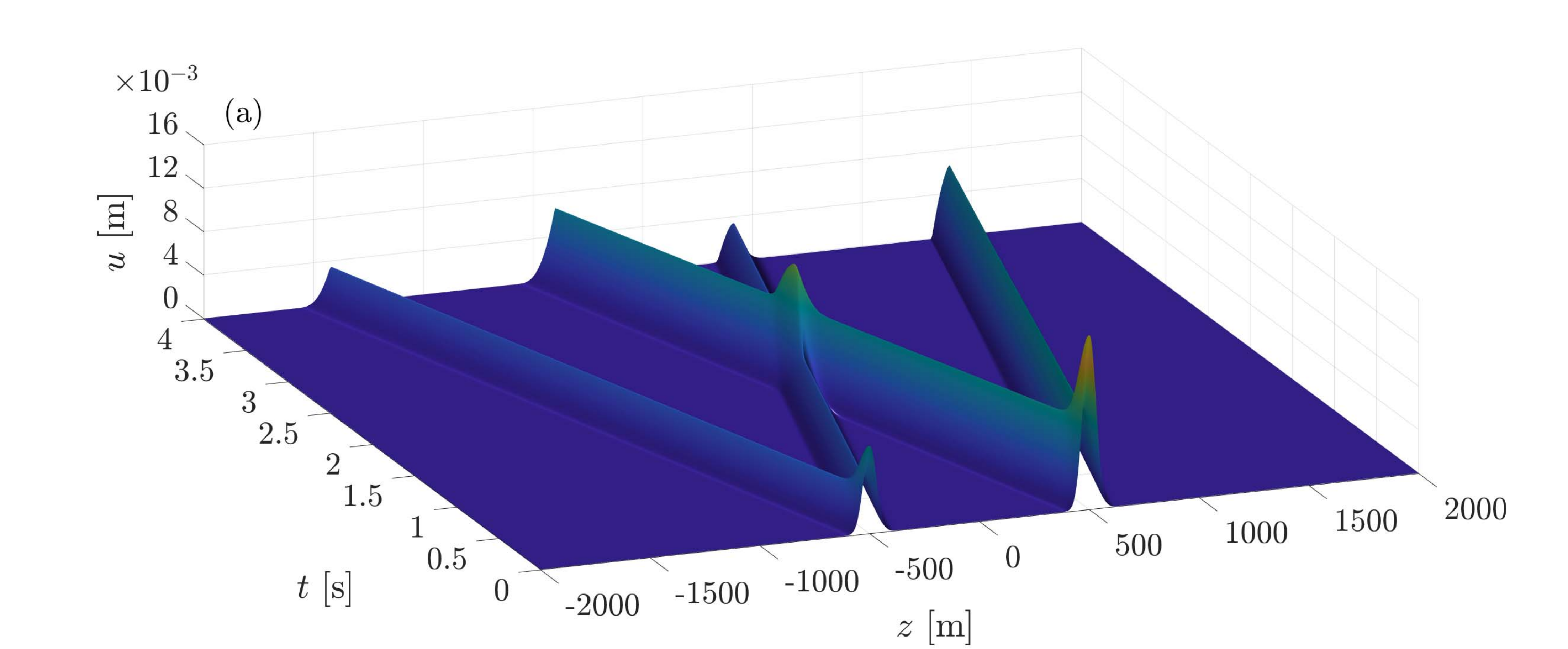}
 	\hspace{-5mm}
 	\includegraphics[scale = 0.28]{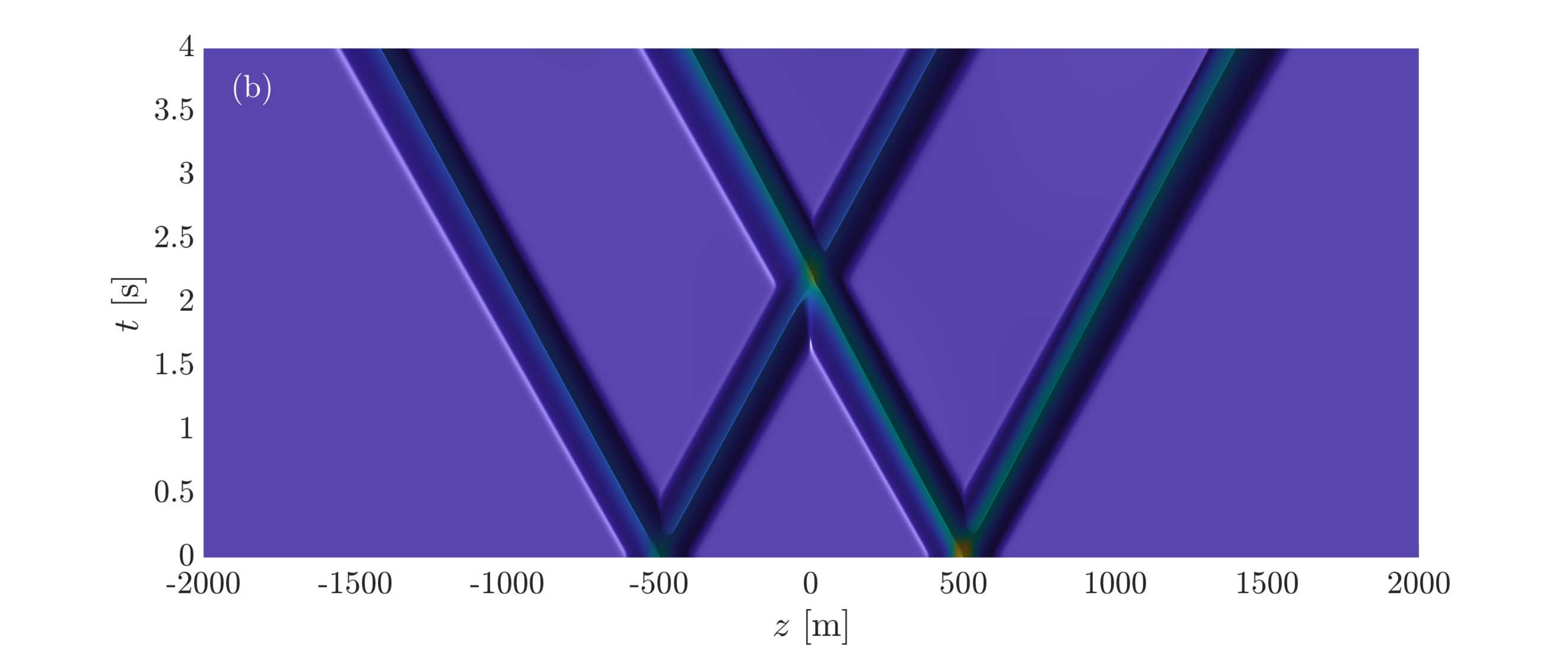}\\
	\includegraphics[scale = 0.36]{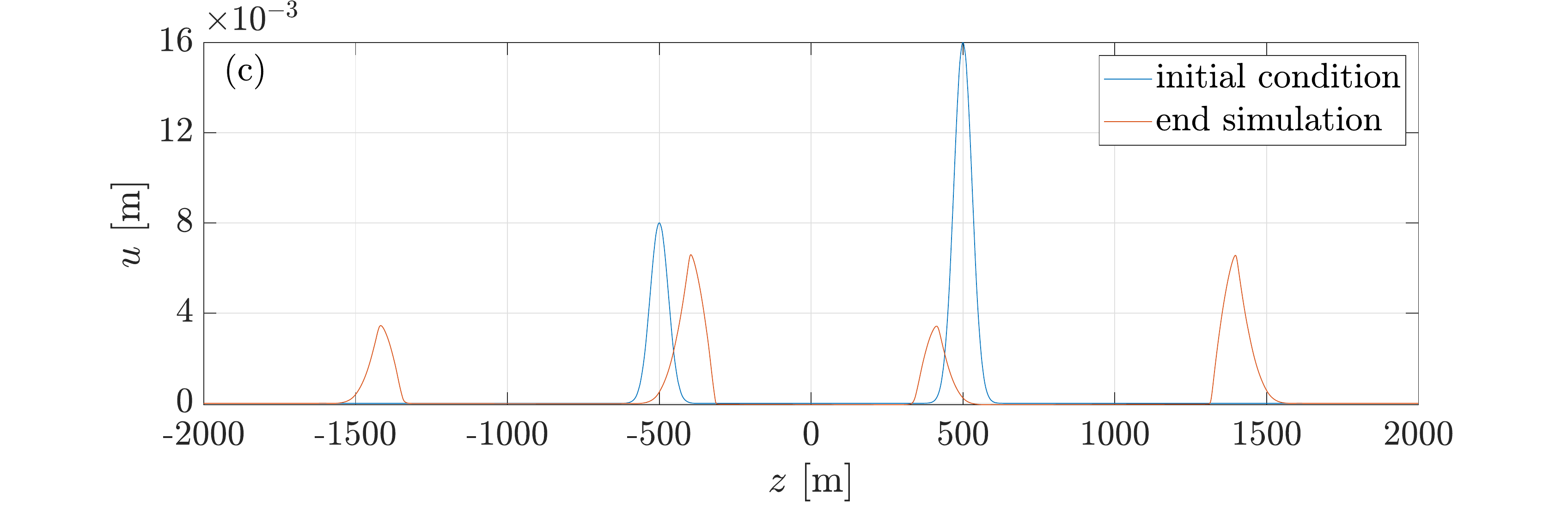}%
    \caption{Interaction of two Gaussian pulses with maximal amplitudes $u_1=0.008\,\textup{m}$ and $u_2=0.016\,\textup{m}$, starting at the positions $z_1=-10\,\textup{m}$ and $z_2=10\,\textup{m}$ for the nonlinear Eq.~\eqref{eq:nonlinEoM2}. The solution is shown from two different perspectives ((a) and (b)) and at the end of the simulation time (c).}
	\label{fig:InteractionPulses}%
\end{figure}

In order to study the interaction of two solutions, two Gaussian pulses starting at $z_1=-500\,\textup{m}$, $z_2=500\,\textup{m}$ with amplitudes ${u_1=0.008\,\textup{m}}$, {$u_2=0.016\,\textup{m}$}, respectively, and the same standard deviation $\sigma=30\, \textup{m}$ are considered. The corresponding initial condition has been calculated by adding both pulses. The obtained temporal evolution of the corresponding solution is displayed in Fig.~\ref{fig:InteractionPulses}. It is shown that both pulses divide directly in two parts, propagating in different directions. Thereby, the interaction of the resulting waves does not destroy the structure of each wave but does lead to a very small change in their amplitude, which can be verified by comparing the waves that did interact with the ones that did not. It can thus be concluded that such waves stay localized after collision with each other. At the same time, these waves are not stationary and their dispersion can be clearly seen in Fig.~\ref{fig:InteractionPulses}, so the waves are in principle not even expected to regain their shape after interaction. We therefore study results for stationary solutions in the next sections.

\subsection{Results for stationary wave solutions}\label{sec:StatWaveSols_results}
In order to obtain the stationary wave solutions, we solve Eq.~\eqref{eq:SolitonSpecificIntegrated}
numerically for $y=u_{,\xi}$. Thereby, the traveling velocity $c$ can be arbitrarily chosen. An integration with respect to $\xi$ yields finally the solution $u$, which is used as initial condition for Eq.~\eqref{eq:nonlinEoM2}.
Since the scheme described in Section~\ref{sec:Scheme} needs also an initial condition at the time point $t^{-1}$, the corresponding values at this time have to be computed as well. Because the stationary solution propagates with the velocity $c$, the solution $u^{-1}$ at time point $t^{-1}$ can be computed by shifting the stationary solution in space by $c\,t^{-1}$.

\begin{figure}[h!]
	\centering	
	\includegraphics[scale = 0.33]{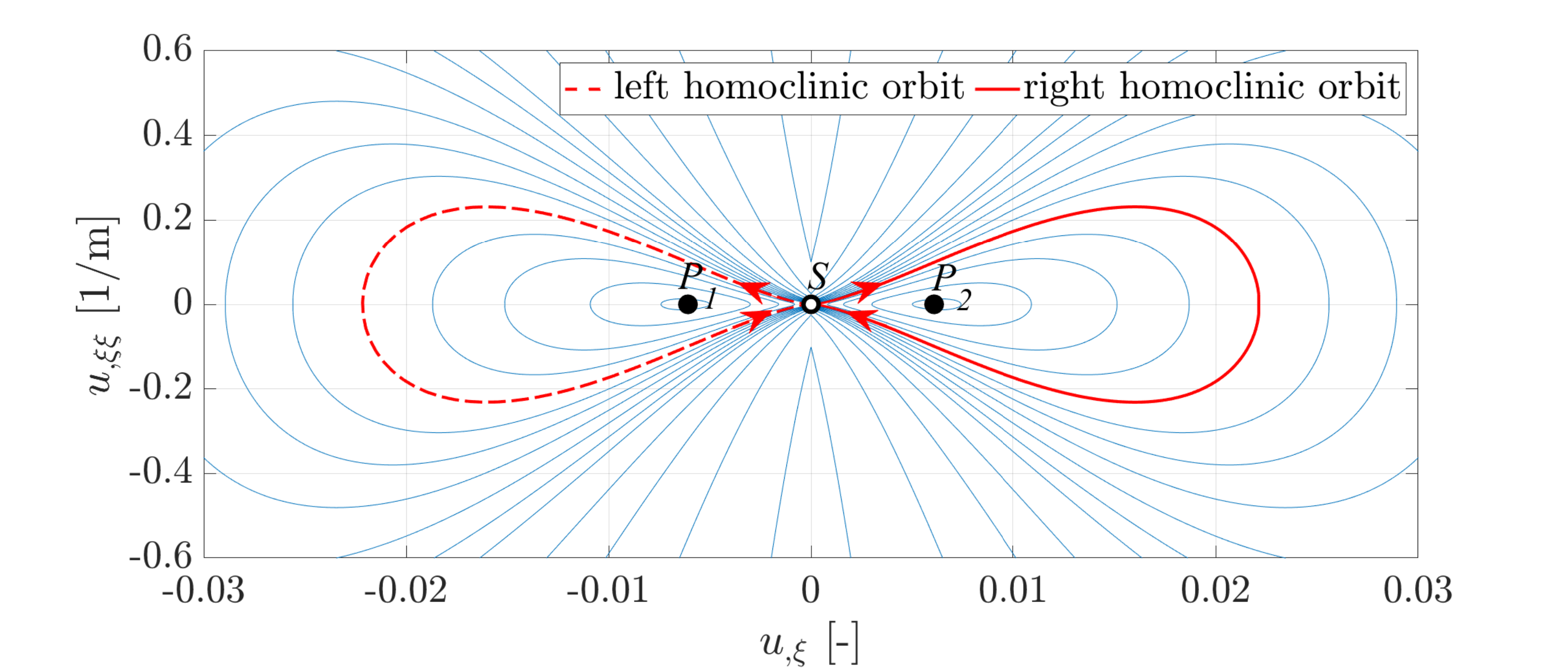}
	\caption{Phase portrait and homoclinic orbits for $c=100\,\textup{m/s}$.}
	\label{fig:homoclinicOrbitPhase}%
\end{figure}

From Eq.~\eqref{eq:SolitonSpecificIntegrated}, we obtain the corresponding phase portrait for $u_{,\xi}$ and $u_{,\xi\xi}$ as shown in Fig.~\ref{fig:homoclinicOrbitPhase} for $c = 100$ m/s, where $P_1$ and $P_2$ denote the two existing fixed points and $S$ denotes the saddle point. Starting from this saddle point, we can find two homoclinic orbits, which we refer to as left homoclinic orbit and right homoclinic orbit. 

\begin{figure}[h!]
	\centering	
	\includegraphics[scale = 0.416]{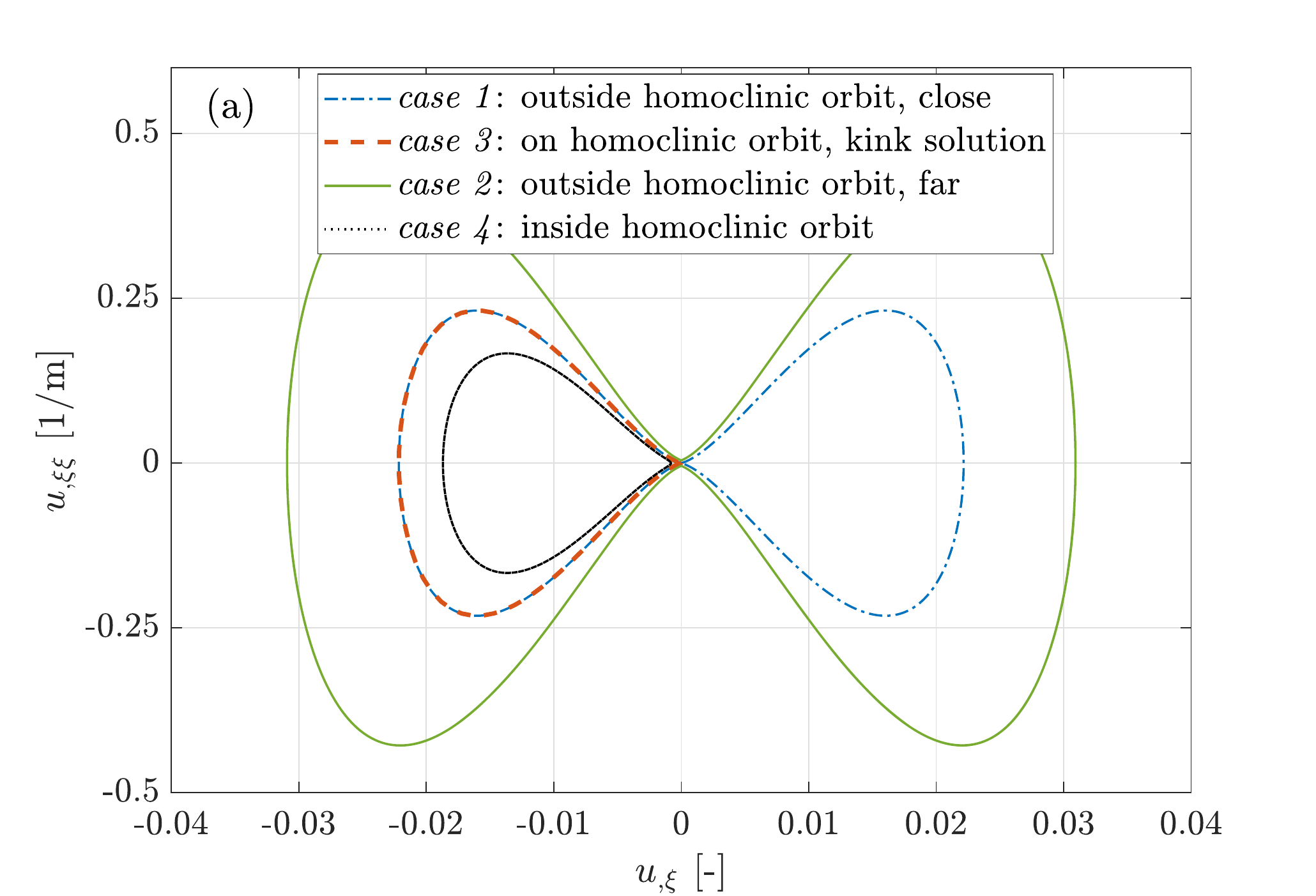}
	\hspace{-5mm}
	\includegraphics[scale = 0.42]{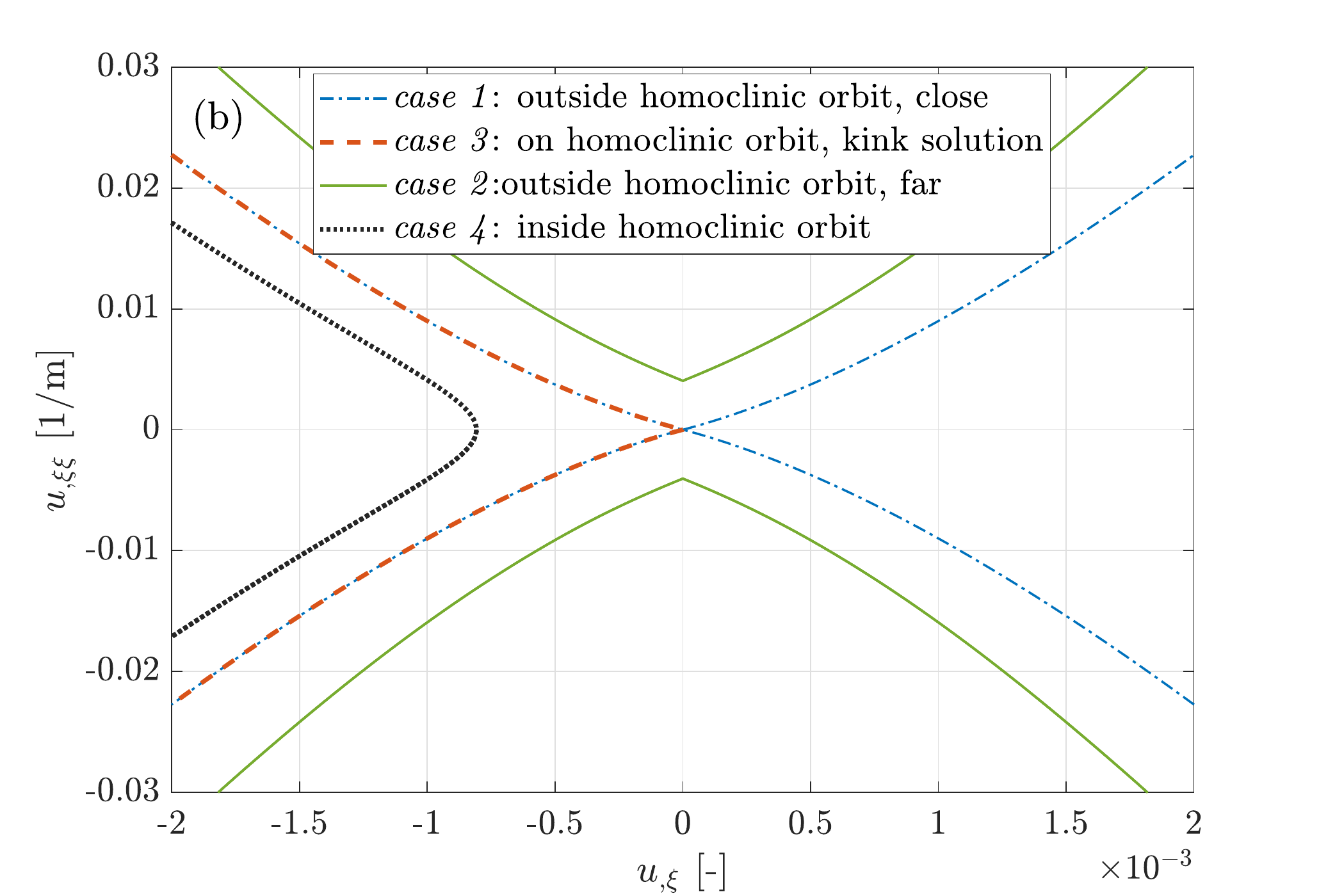}\\
	\caption{a: Phase plot of the trajectories of \textit{cases 1-4}.
	b: Zoomed phase plot of the trajectories of \textit{cases 1-4} close to the origin of the phase space.}%
	\label{fig:phase portrait_cases}%
\end{figure}

In the following investigation of stationary solutions, we look at four cases:
\begin{itemize}
    \item Stationary solution, which is located in the phase portrait “outside homoclinic orbit, close” (in the following denoted as \textit{case 1}).
    \item 	Stationary solution, which is located in the phase portrait “outside homoclinic orbit, far” (in the following denoted as \textit{case 2}).
    \item Stationary solution, which is located on the left homoclinic orbit (in the following denoted as \textit{case 3}).
    \item Stationary solution, which is located in the phase portrait “inside homoclinic orbit” (in the following denoted as \textit{case 4}).
\end{itemize}
The four corresponding trajectories in the phase portrait are shown in Fig.~\ref{fig:phase portrait_cases}a. Thereby, the blue colored trajectory corresponds to \textit{case 1}, the black colored trajectory to \textit{case 2}, the dashed red colored trajectory to \textit{case 3} and the yellow colored trajectory \textit{case 4}. Fig.~\ref{fig:phase portrait_cases}b shows the behavior of the trajectories close to the origin of the phase portrait. The trajectories of \textit{case 1} and \textit{case 3} do not lie on top of each other. \textcolor{black}{In \textit{case 1} the trajectory encircles both homoclinic orbits, whereas in \textit{case 3} it is located on the left homoclinic orbit.}

\begin{figure}[h!]
	\centering	
	\includegraphics[scale = 0.28]{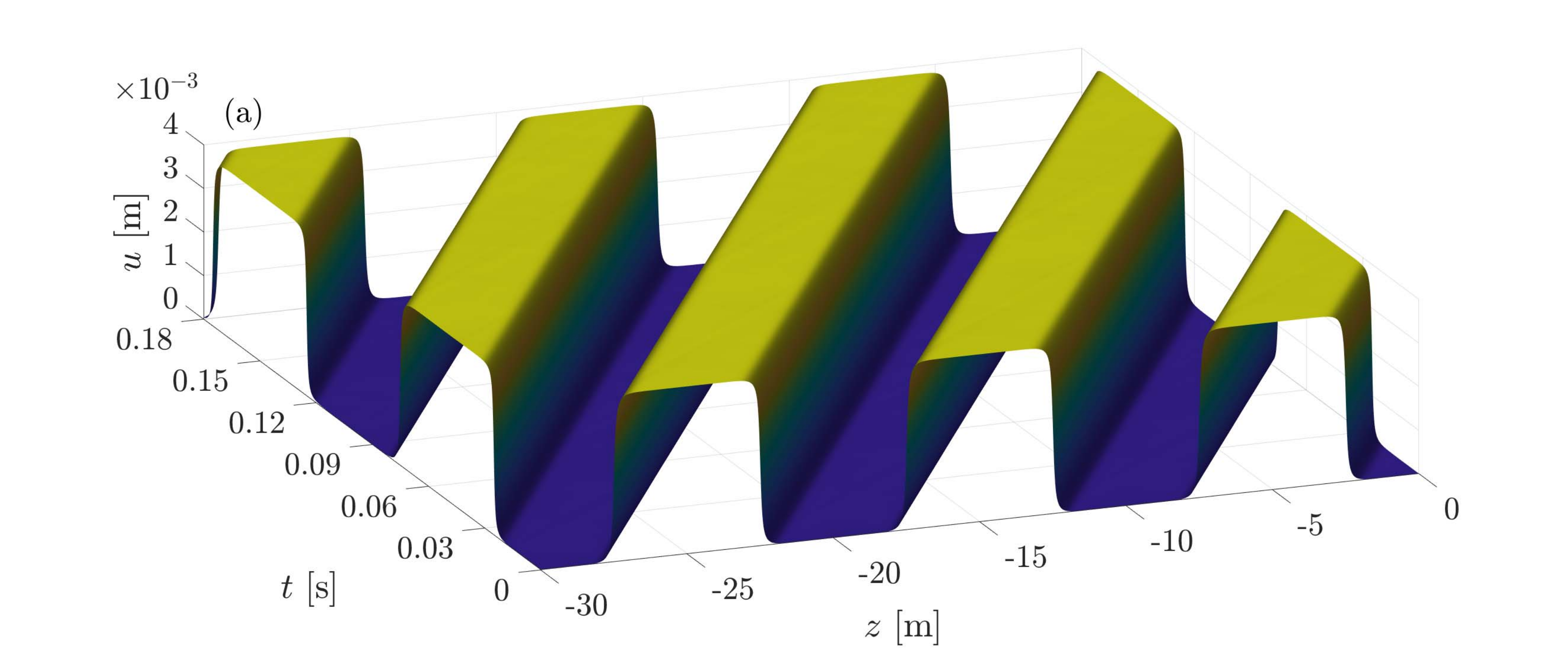}
	\hspace{-5mm}
	\includegraphics[scale = 0.32]{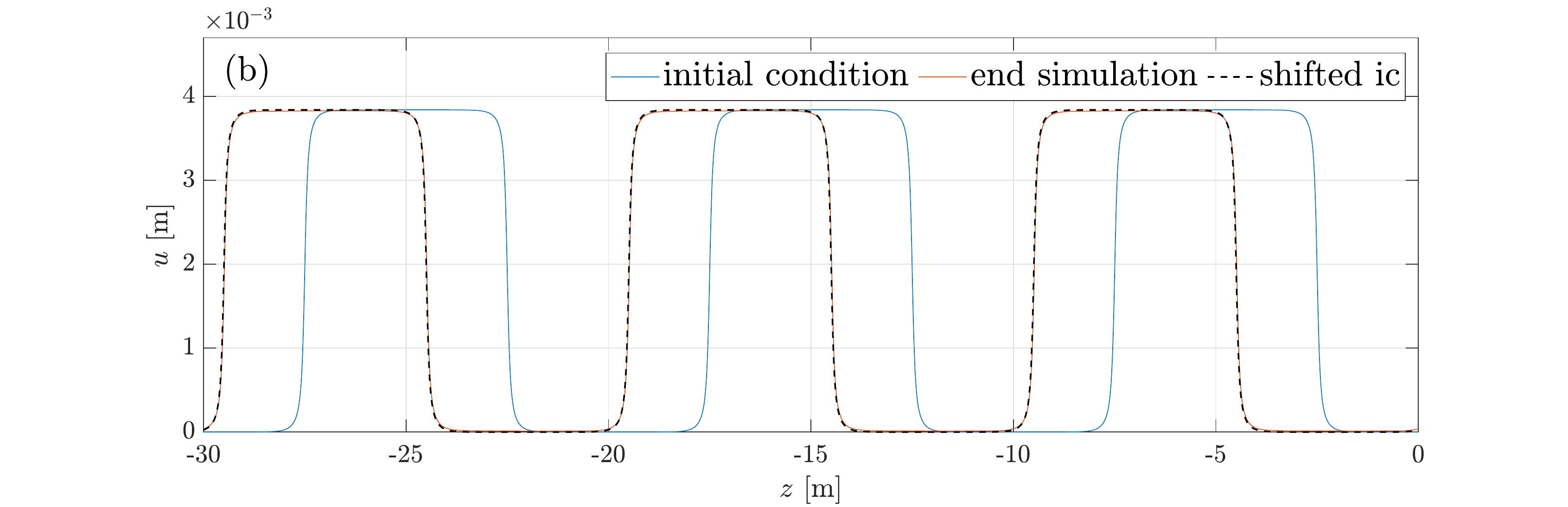}\\
	\includegraphics[scale = 0.32]{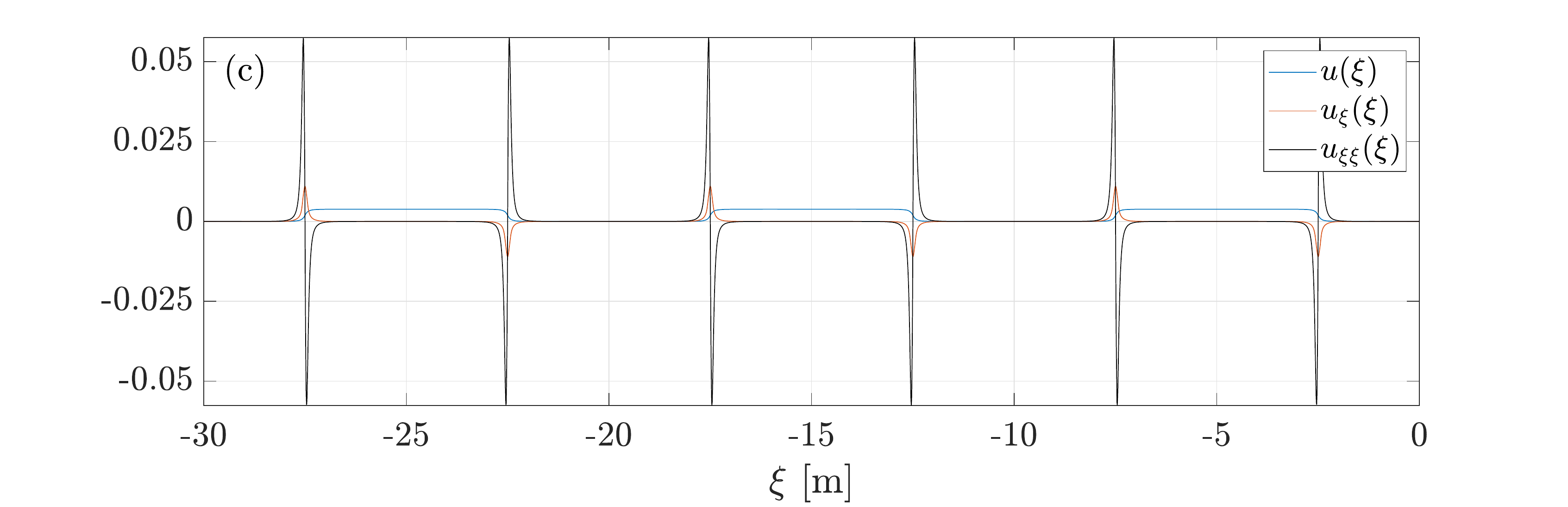}\\
	\caption{a: Temporal evolution, b: end time simulation results, and c: corresponding evolution of $u(\xi)$, $u_{,\xi}(\xi)$, and $u_{,\xi\xi}(\xi)$  for \textit{case 1}.}%
	\label{fig:simulation_case1}%
\end{figure}

\begin{figure}[h!]
	\centering	
	\includegraphics[scale = 0.28]{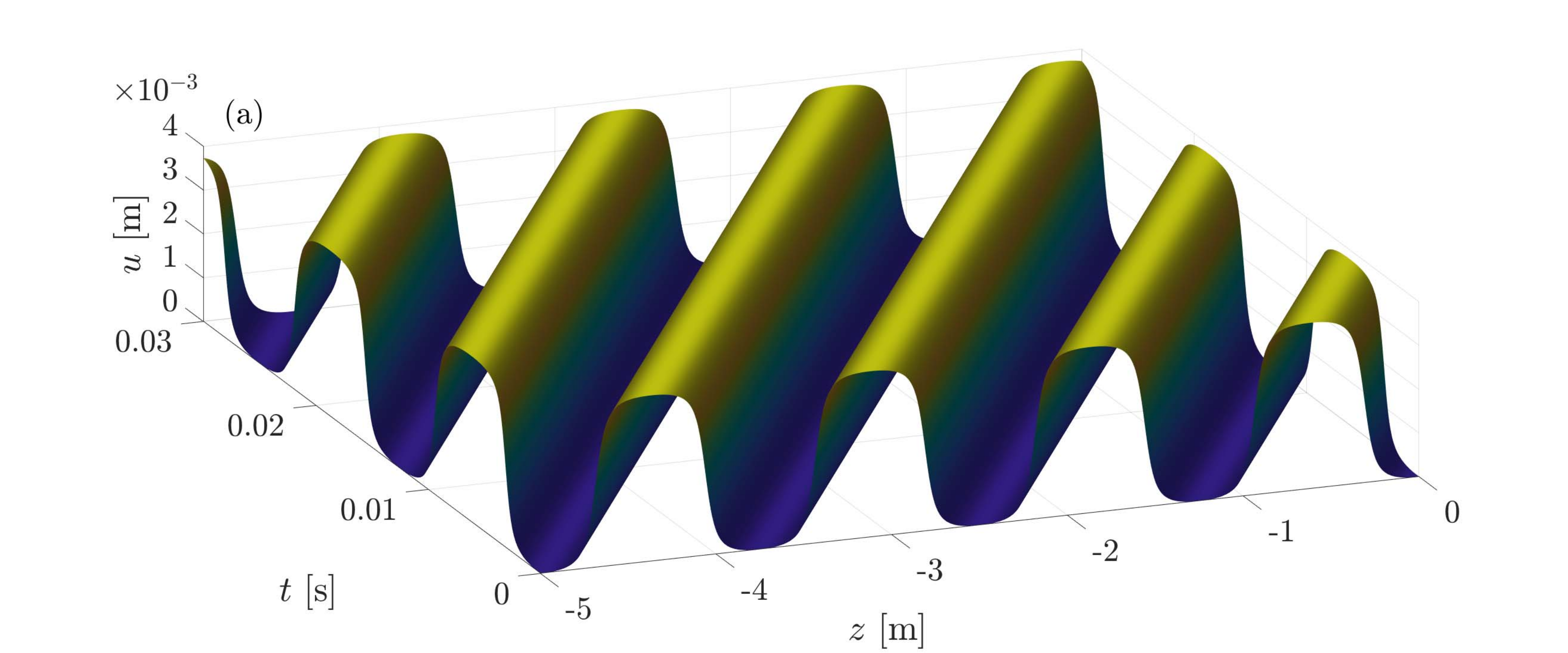}
	\hspace{-5mm}
	\includegraphics[scale = 0.32]{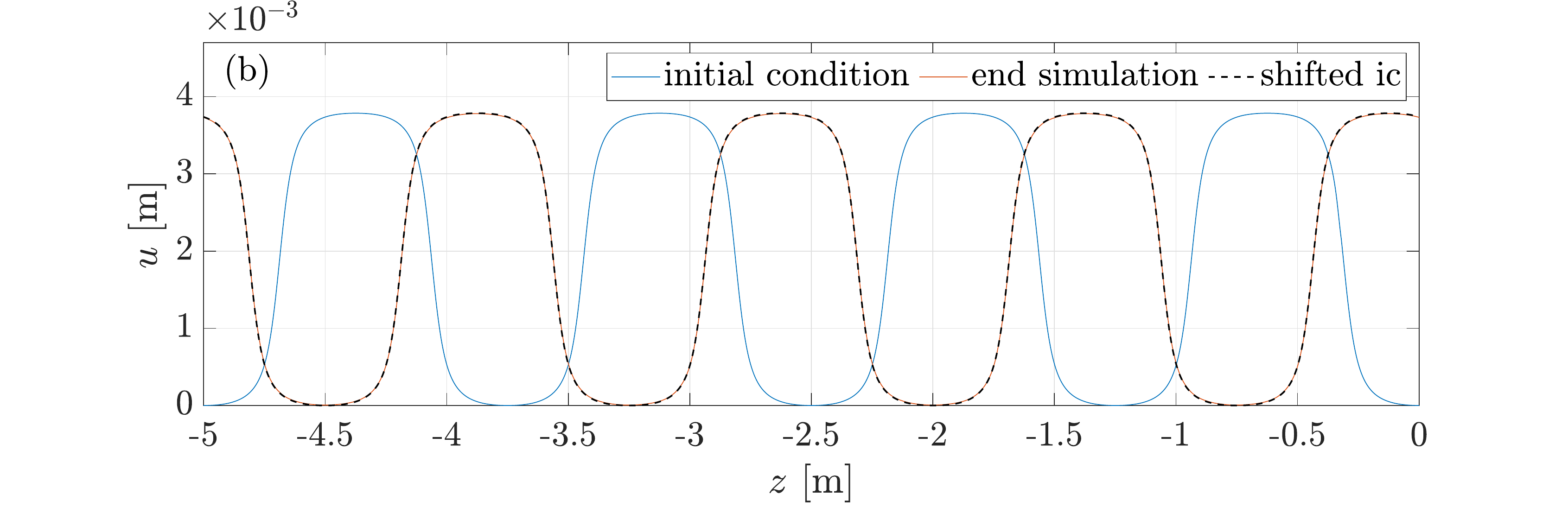}\\
	\includegraphics[scale = 0.32]{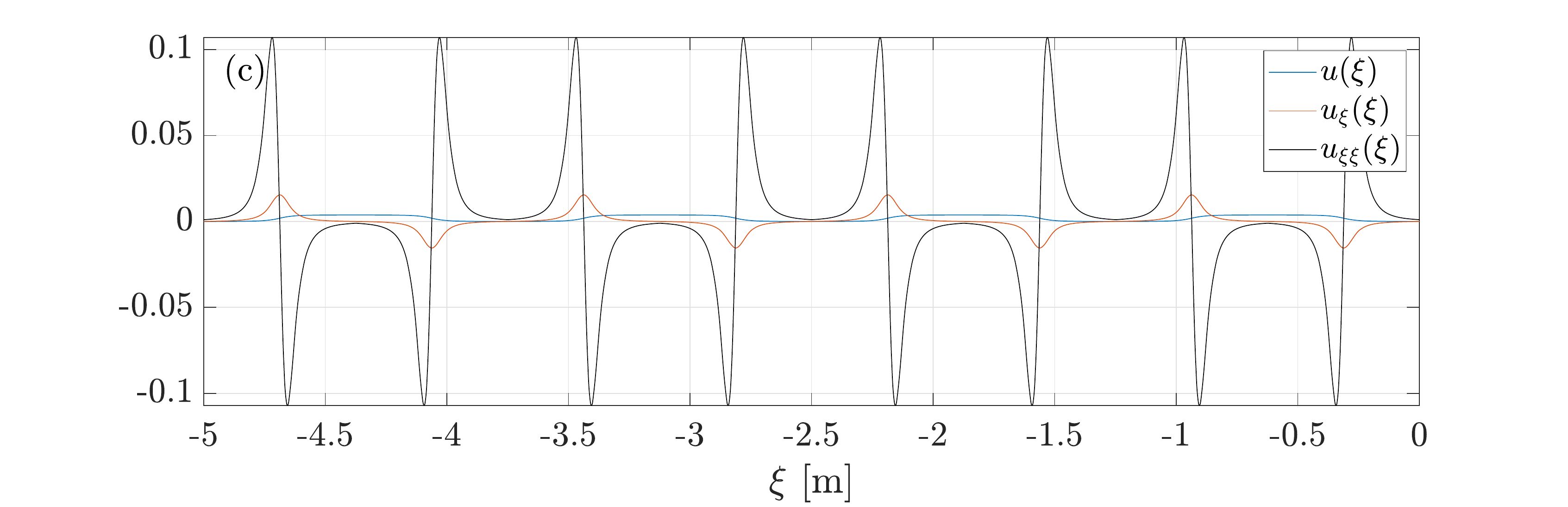}\\
	\caption{a: Temporal evolution, b: end time simulation results, and c: corresponding evolution of $u(\xi)$, $u_{,\xi}(\xi)$, and $u_{,\xi\xi}(\xi)$  for \textit{case 2}.}%
	\label{fig:simulation_case2}%
\end{figure}

\begin{figure}[h!]
	\centering	
	\includegraphics[scale = 0.28]{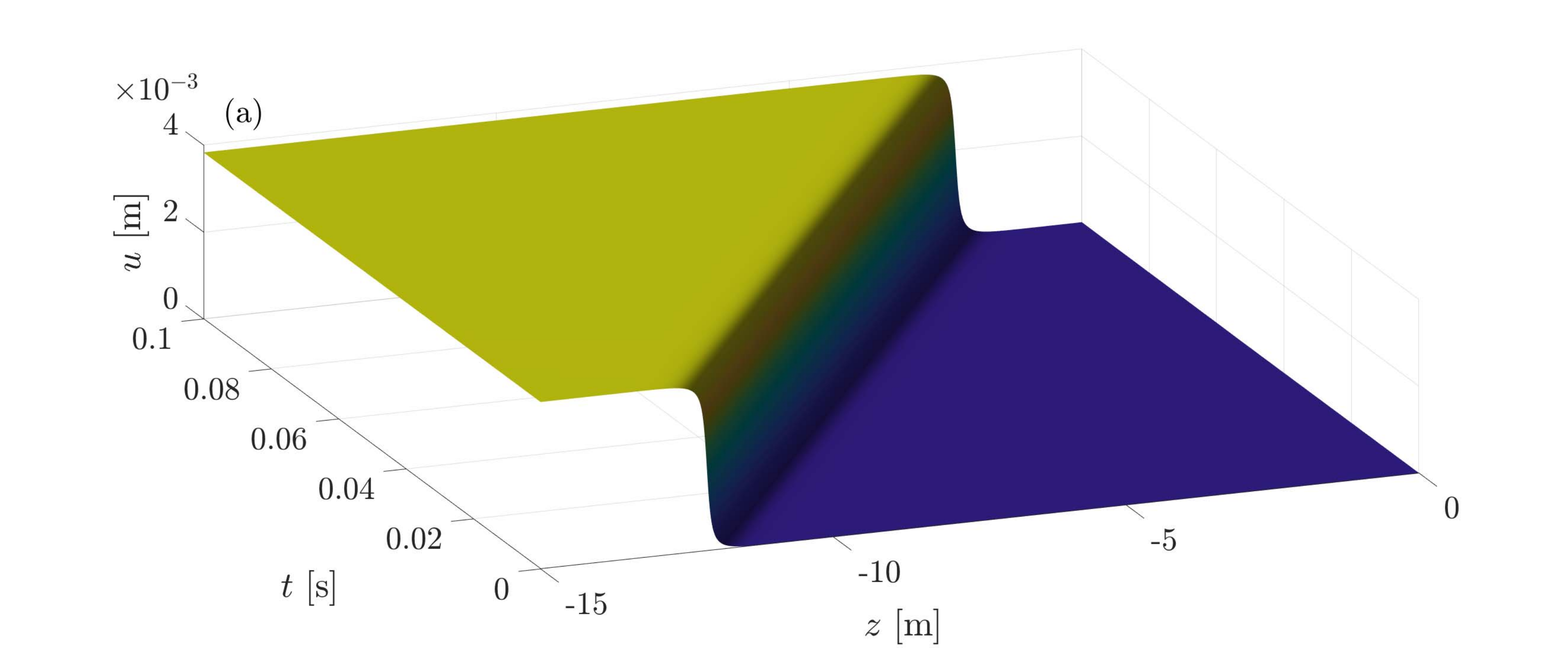}
	\hspace{-5mm}
	\includegraphics[scale = 0.32]{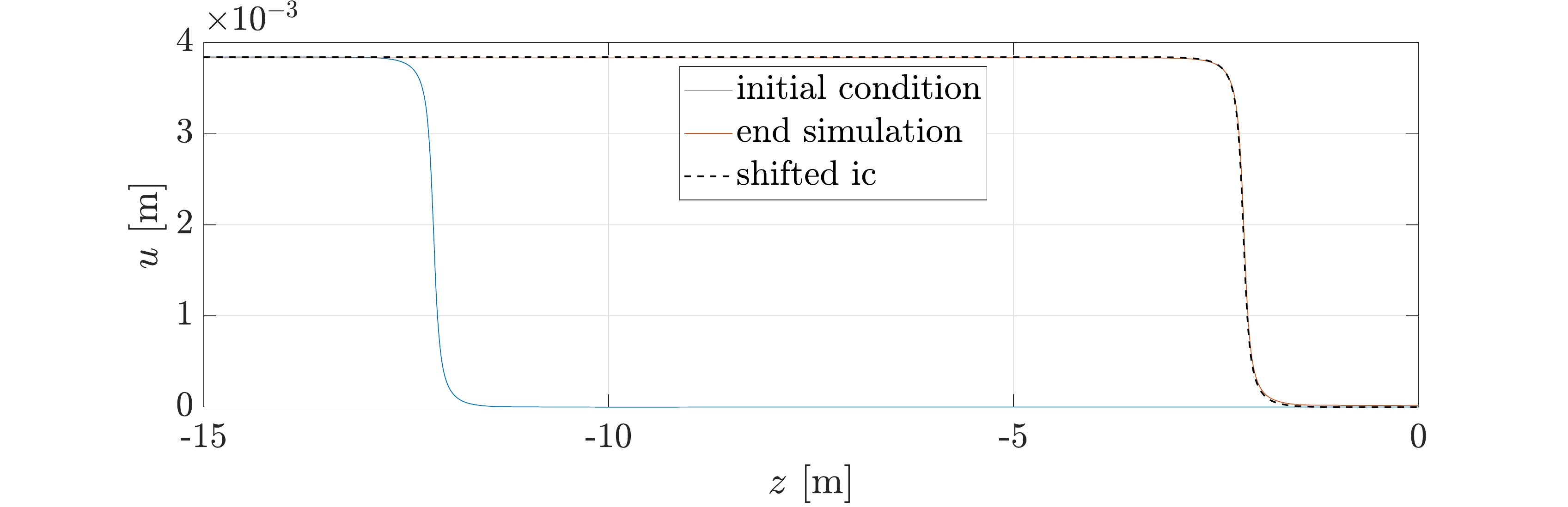}\\
		\includegraphics[scale = 0.32]{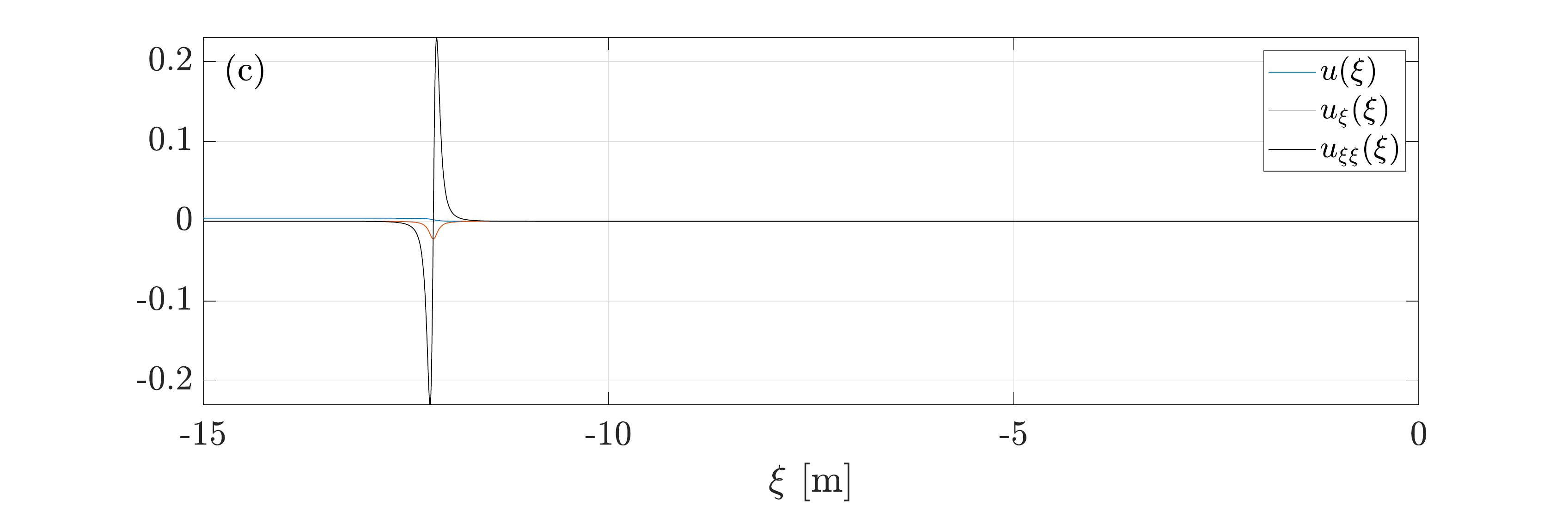}\\
	\caption{a: Temporal evolution, b: end time simulation results, and c: corresponding evolution of $u(\xi)$, $u_{,\xi}(\xi)$, and $u_{,\xi\xi}(\xi)$  for \textit{case 3}.}%
	\label{fig:simulation_case3}%
\end{figure}

\begin{figure}[h!]
	\centering	
	\includegraphics[scale = 0.28]{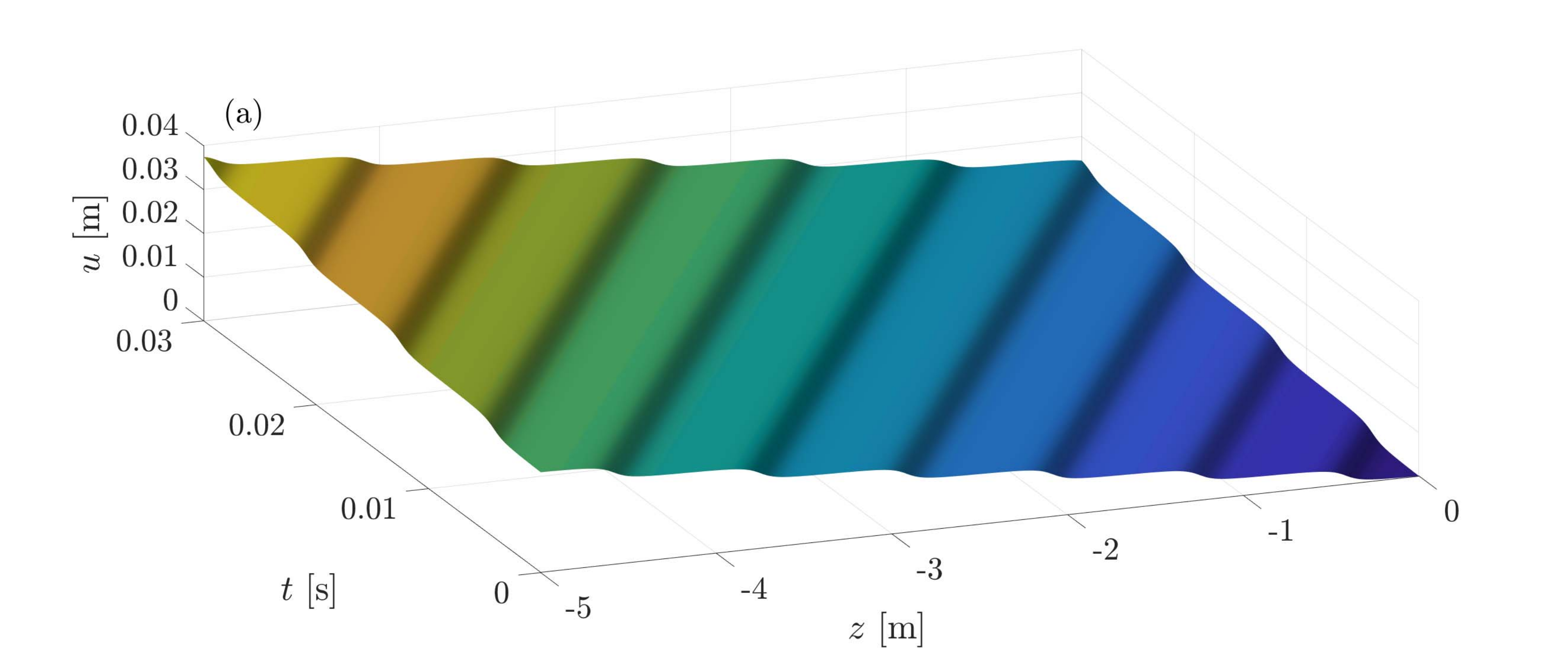}
	\hspace{-5mm}
	\includegraphics[scale = 0.32]{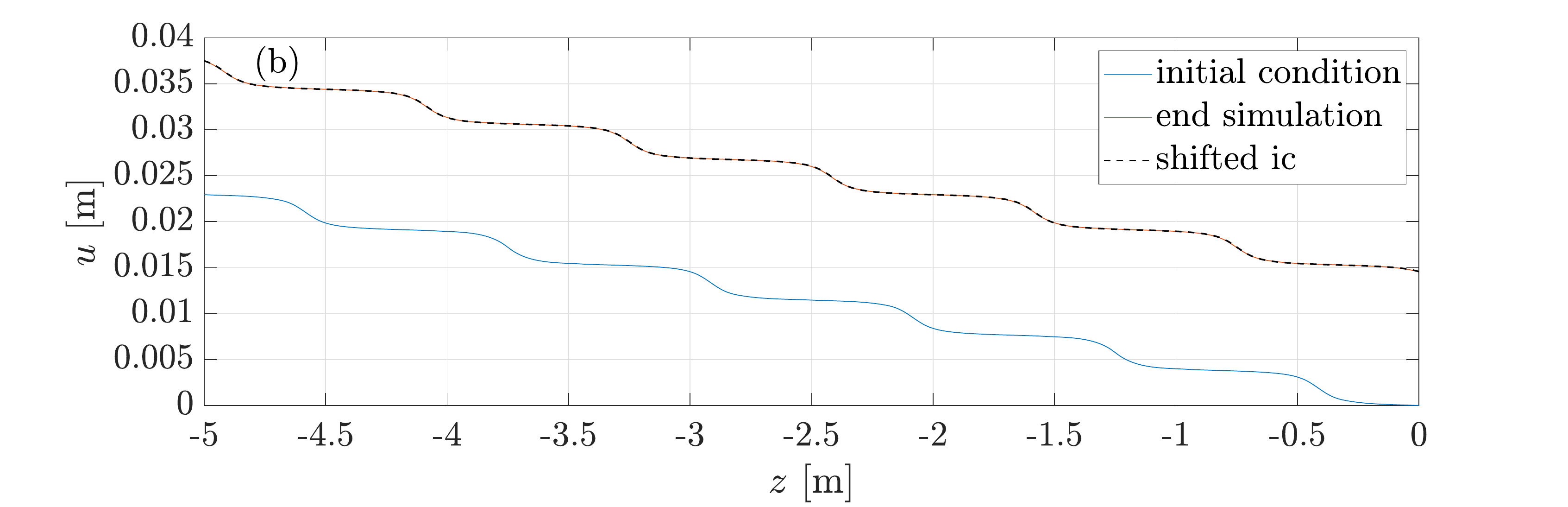}\\
	\includegraphics[scale = 0.32]{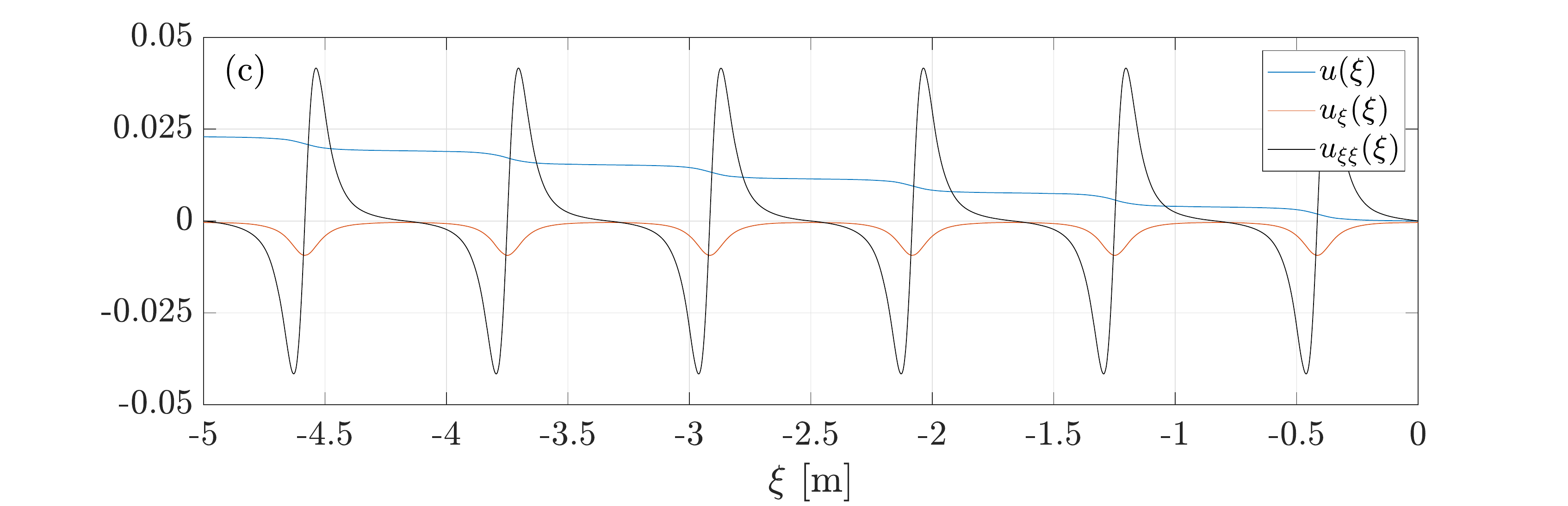}\\
	\caption{a: Temporal evolution, b: end time simulation results, and c: corresponding evolution of $u(\xi)$, $u_{,\xi}(\xi)$, and $u_{,\xi\xi}(\xi)$  for \textit{case 4}.}%
	\label{fig:simulation_case4}%
\end{figure}

Next, we show the temporal evolution of the four solutions related to \mbox{\textit{cases 1-4}} in detail. The corresponding results are shown in Figs.~\ref{fig:simulation_case1}-\ref{fig:simulation_case4}. In each figure, the temporal evolution of the solution is shown and the initial condition is compared to the solution at the end of the simulation. In each computation, the velocity of the waves is chosen as $c=100$ m/s. 
For the computation of each solution, periodic boundary conditions have been used.
In these results, we can identify two periodic solutions, which correspond to \textit{case 1} and \mbox{\textit{case 2}}. Clearly, the periodic pattern converges to a more edged pattern as the periodic solution gets closer to the homoclinic orbits.
Starting at one of the homoclinic orbits, we obtain the so-called \textit{kink solution} or \textit{kink~wave}, which also appears on a homoclinic orbit for the sine-Gordon equation~\cite{ablowitz1973method,ivancevic2013sine}, see appendix~\ref{Appendix:sineGordon}. This solution is not periodic anymore and stays constant before and after the kink in solution.
Another solution is obtained when starting inside a homoclinic orbit, like in \mbox{\textit{case 4}}. Then, we obtain a descending periodic pattern in the stationary solution if starting inside the left homoclinic orbit and an ascending periodic pattern, if starting inside the right homoclinic orbit.
For \textit{case~4}, modified periodic boundary conditions have been used in order to periodically continue the pattern at the boundaries. For this case, the derivatives of the solution at the left and right end of the spatial computation domain are equal and the difference between the corresponding function values of the solution are constant in time. Based on this, the difference at the right end of the spatial domain is added such that there is a periodic transition from the right to the left end value of the solution.

\textcolor{black}{The amplitude of the stationary waves $u(\xi)$ can be controlled using the parameter $c$. In addition, the width of the stationary waves depends on the distance to a homoclinic orbit. The closer the phase-portrait trajectories ($u_{,\xi},u_{,\xi\xi}$) are to a homoclinic orbit, as shown in Fig.~\ref{fig:homoclinicOrbitPhase}, the wider the stationary waves $u(\xi)$ are.}

As can be seen in all figures, the form of the used initial condition is maintained in the corresponding temporal evolution. In other words, different parts of the solution do not interact with each other and stay stationary, which is to be expected. This implies that such waves in reality may be able to propagate over large distances and, when excited by an earthquake, even reach the surface. This is particularly relevant for the kink wave, which may be excited due to a sudden permanent local displacement associated with sliding along a fault.





\subsection{Interaction of kink solutions}\label{sec:KinkInteraction}

\begin{figure}[h!]
	\centering	
	\includegraphics[scale = 0.28]{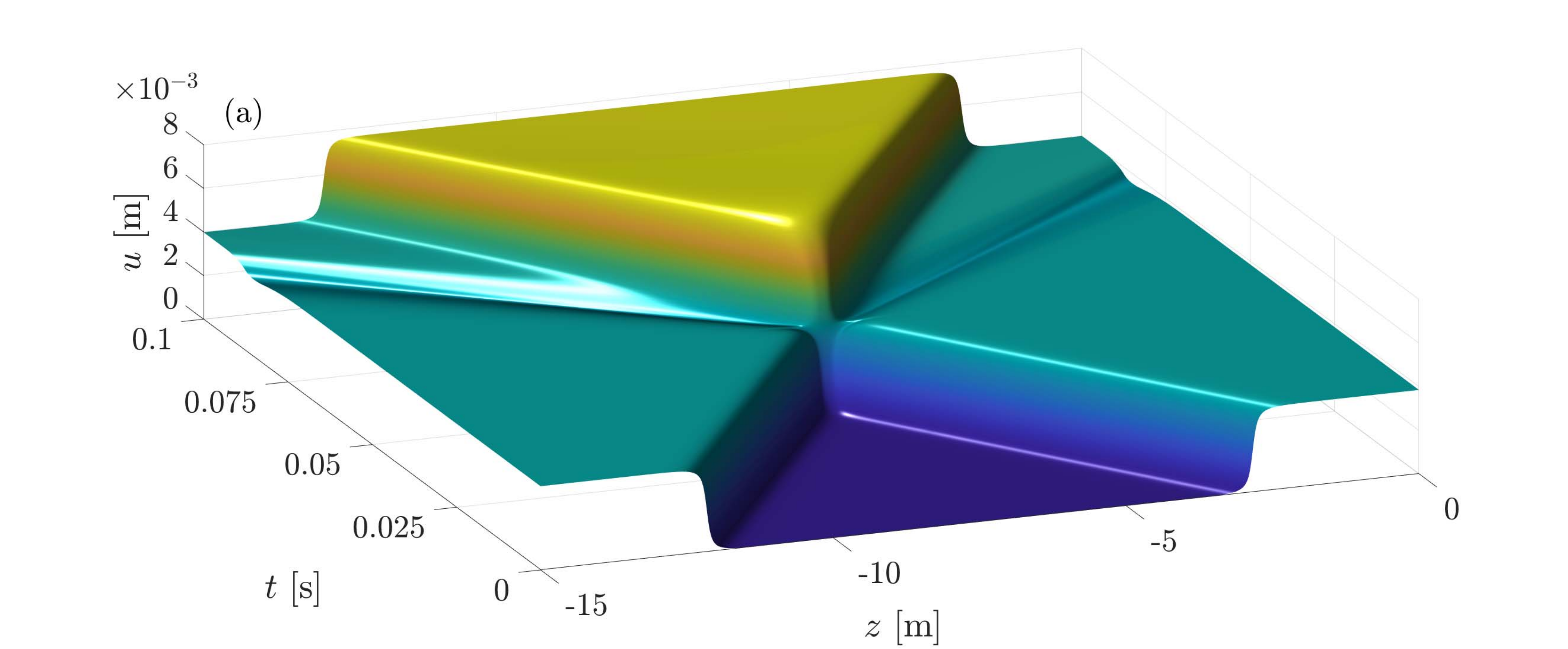}
	\includegraphics[scale = 0.31]{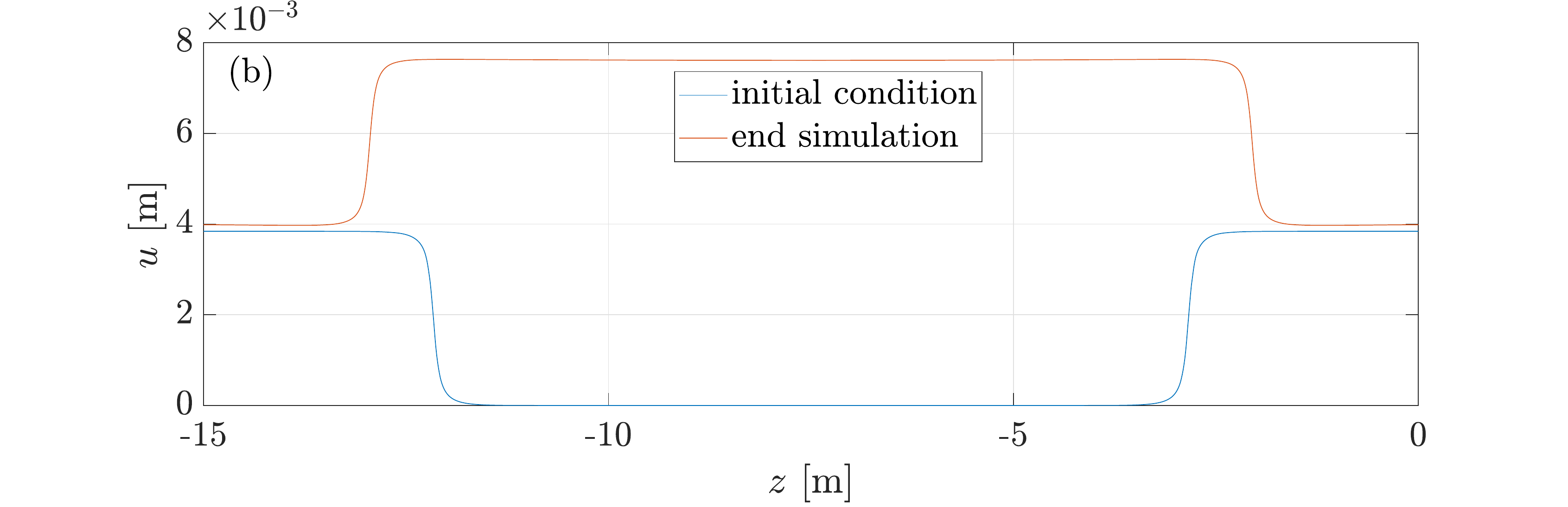}
	\includegraphics[scale = 0.31]{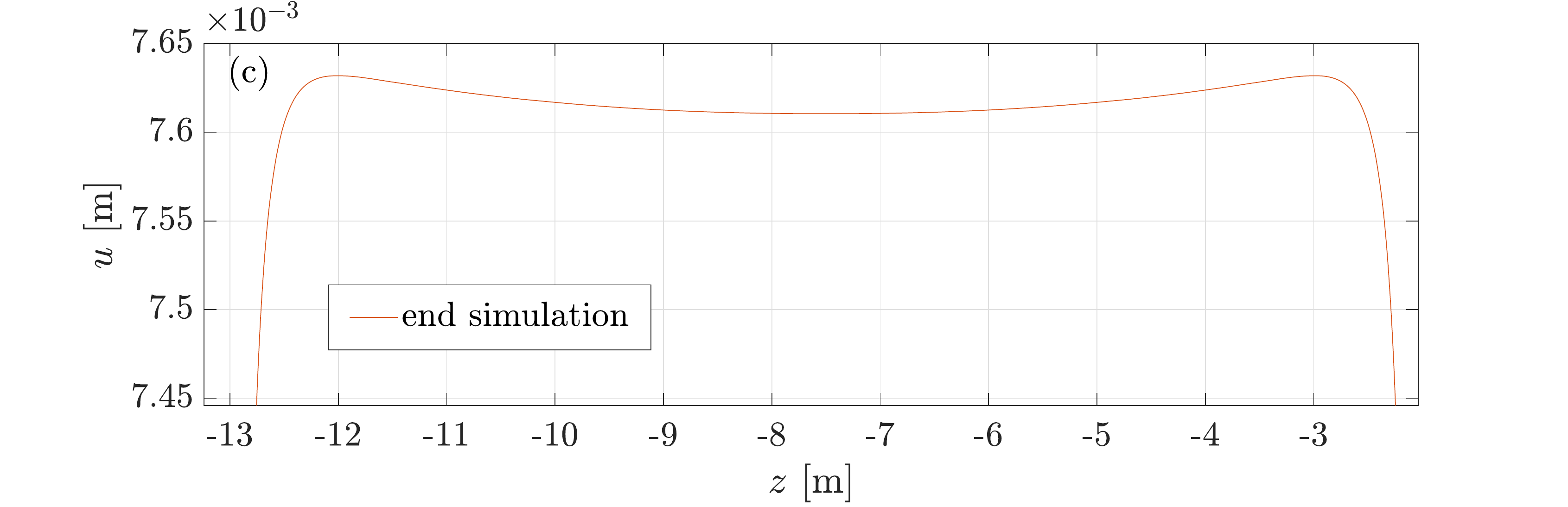}
	\caption{a: Interaction of two kink wave solutions. b: the corresponding initial condition as well as the solution at the end of the simulation. c: Zoomed plot of the solution at the end of the simulation. These results are obtained for $c=100\,\textup{m/s}$.}
	\label{fig:KinkSoliton_Collision_c100}%
\end{figure}

%
The result of interaction of two kink wave solutions is shown in Fig.~\ref{fig:KinkSoliton_Collision_c100}, for \mbox{$c=100$ m/s}. \textcolor{black}{We observe an interesting nonlinear behaviour, which is different from classical soliton collision experiments where localized solutions regain their complete shape after collision, as for example is the case for the Korteweg-de Vries equation ~\cite[e.g.,][]{newell1985solitons}. Compared to the initial condition, the displacement amplitude at the end of the simulation has become slightly larger at the left and right sides of the considered spatial domain; the displacement has the value of $3.98 \text{ mm}$ compared to $3.84 \text{ mm}$ for the initial condition. Moreover, as shown in Fig.~\ref{fig:KinkSoliton_Collision_c100}c, the plateau at the end of the simulation is slightly bent downwards compared to the initial condition, which is flat (i.e., zero) in the spatial interval from $z=-9\text{ m}$ to $z=-6\text{ m}$. This difference compared to the corresponding values of the initial condition (i.e., stationary waves) has been analyzed for different spatial and temporal discretizations; for each of the considered discretizations, the difference did not vanish.
Therefore, we can conclude that the kink solutions change their shape due to interaction. Hence, the kink wave cannot be referred to as a true soliton.}


\section{Discussion}\label{sec:Discussion}

In this section, we discuss two aspects regarding the model adopted in this paper: a potential improvement to overcome the presence of evanescent waves and a potential way to obtain closed-form expressions for the additional elastic moduli in the equation of motion. Regarding the first point, it can be verified that the linear version of our model is stable for all possible initial conditions (i.e., for all wavenumbers) and  that the group velocity associated with the propagating wave is finite for all wavenumbers. However, the model does allow the presence of an evanescent wave through which information can travel at infinite speed. This is a short-coming of the model as it does not comply with the so-called strict/Einstein's causality condition~\cite{metrikine2006causality}. It is expected that the nonlinear gradient elasticity model has this short-coming too. For the linear problem, it can be overcome by adding a term with fourth-order time derivative in the equation of motion so as to include the effect of micro-scale inertia (thus balancing the highest order of spatial and temporal derivatives); then, all waves appear to have finite propagation speed~\cite{metrikine2006causality}. Such a term is often included in the equation of motion of nonlinear gradient elasticity models too~\cite{berezovski2013dispersive}.

Secondly, we recall that the equation of motion of the nonlinear gradient elasticity model has been obtained using an assumed strain dependence of the additional elastic moduli (Eq.~\eqref{eq:elasticParametersG}). Generally speaking, the equation of a higher-order gradient elasticity model can be derived starting from a periodically layered classical continuum and applying an asymptotic homogenization procedure~\cite{fish2002non,andrianov2008higher}. By applying such a procedure for a nonlinear periodically layered classical continuum, closed-form expressions may be found for the additional elastic moduli (of the homogenized medium) in terms of the strain-dependent shear moduli of the original layers and their densities. It may also reveal how the time and length scales should be chosen, and with that the coefficients $B_1$ and $B_2$. Finally, it may naturally lead to the inclusion of the fourth-order time-derivative term representing the effect of micro-scale inertia referred to above.

\section{Conclusions}\label{sec:Conclusions}
The aim of this paper was to investigate the existence of localized stationary waves in the shallow subsurface with constitutive behaviour governed by the hyperbolic model, implying that the shear modulus is strain dependent (i.e., non-polynomial nonlinearity). To this end, we derived a novel equation of motion for a nonlinear gradient elasticity model, which is of the Boussinesq-type. The higher-order gradient terms (compared to those in the classical wave equation) capture the effect of small-scale soil heterogeneity/micro-structure, which introduces dispersive effects particularly for the shorter waves; the dispersion prohibits the formation of jumps, which leads to physically realisable solutions. In order to obtain stationary solutions of the equation of motion, we introduced the moving reference frame together with the stationarity assumption, which yields an ordinary differential equation. We also presented a novel numerical scheme to solve the nonlinear equation of motion in space and time \textcolor{black}{up to an order of accuracy of $\mathcal{O}(\Delta t^2+\Delta z^2)$}, which exploits the structure of the partial differential equation in order to simplify the \textcolor{black}{computation} of the spatial finite-difference approximations.

Periodic (with and without a descending trend) as well as localized stationary waves were found by analyzing the above mentioned ordinary differential equation in the phase portrait, and integrating it along the different trajectories. The localized stationary wave is in fact a kink wave and was obtained by integration along a homoclinic orbit. Generally speaking, the closer the trajectory lies to a homoclinic orbit, the sharper the edges of the corresponding periodic stationary wave and the larger its width (i.e., period). The numerical scheme was used to verify the stationary character of the waves. In addition, it was used to study the propagation of arbitrary initial pulses, which clearly reveals the influence of the nonlinearity: strain-dependent speed in general and, as a result, sharpening of the pulses. Finally, we simulated using the numerical scheme a collision experiment in which two kink waves propagate in opposite direction and pass each other. We found that, after interaction, their original shapes are not recovered. Therefore, we conclude that the kink wave identified in this work is in fact not a true soliton. 

Even though not being a true soliton, the kink wave, which may have high amplitude, can propagate through the soil column and potentially reach the surface depending on the strength of the material and geometrical damping mechanisms (which have not been considered). Therefore, seismic site response analyses should not a priori exclude the presence of such localized stationary waves. Follow-up research should be devoted to quantifying the influence of the damping mechanisms on the decay of the waves for specific soil profiles; here, the Masing model for hysteresis could serve to include the material damping~\cite{lacarbonara2003nonclassical,lacarbonara2012nonlinear}.




\appendix
\

\section{Numerical scheme for the linear equation of motion waves}\label{Appendix:SchemeSeismicWaves_Linear}
In order to solve Eq. (\ref{eq:linEoM1}), which is expressed as
\begin{equation} \label{eq:simplified_lin_eq}
    \rho \, u_{,tt}=G_0 u_{,zz} -B_1 L^2 G_0 u_{,zzzz} + B_2 \rho L^2 u_{,ttzz},
\end{equation}
a finite difference scheme is introduced.
Applying explicit finite difference approximations, Eq.~\eqref{eq:simplified_lin_eq} can be written as
\begin{equation}
    \begin{aligned}
        &\rho \, \frac{u_i^{n+1}-2u_i^{n}+u_i^{n-1}}{\Delta t^2}=\\ &G_0\frac{u_{i+1}^{n}-2u_{i}^{n}+u_{i-1}^{n-1}}{\Delta z^2}-B_1 L^2 G_0 \frac{u_{i+2}^{n}-4u_{i+1}^{n}+6u_{i}^{n}-4u_{i-1}^{n-1}+u_{i-2}^{n}}{\Delta z^4} \\
        &+ B_2 \rho L^2 \frac{u_{i+1}^{n+1}-2u_{i}^{n+1}+u_{i-1}^{n+1}-2u_{i+1}^{n}+4u_{i}^{n}-2u_{i-1}^{n}+u_{i+1}^{n-1}-2u_{i}^{n-1}+u_{i-1}^{n-1}}{\Delta t^2\Delta z^2}.
    \end{aligned}
\end{equation}

Collecting all unknown values of $u$ at the timepoint $t^{n+1}$ at the left-hand side and the rest at the right-hand side leads to a system of linear algebraic equations, which has to be solved for each time step. In order to simulate the solution in an open domain, absorbing boundary conditions have been used.

\section{Stationary solution of the sine-Gordon equation}\label{Appendix:sineGordon}
A well known equation which exhibits the so-called kink solitons is the sine-Gordon equation~\cite{ablowitz1973method,ivancevic2013sine}. 
In this appendix, we demonstrate that such a solution can be obtained by introducing a moving reference frame, assuming stationarity, and integrating the thus obtained equation along a homoclinic orbit in the phase portrait. In this paper, we applied essentially the same procedure to get localized solutions of the equation describing stationary waves in the nonlinear gradient elasticity model (Eq. (18)).

We show in the following relations between kink solutions of the equation of motion of the nonlinear gradient elasticity model (Eq.~\eqref{eq:nonlinEoM2}) and kink solutions of the sine-Gordon equation.  

Since it was realized that the sine-Gordon equation led to kink solitons, the importance of this equation greatly increased. The sine-Gordon equation has several important physical applications~\cite{josephson1962possible,gl1971analytical,kjems1978evidence,dodd1982soliton}. 
It can be stated as
\begin{equation}\label{eq:sineGordon}
u_{,tt}-u_{,zz}+\sin u=0,
\end{equation}
where $z\in\mathbb{R}$ denotes the space variable, $t\in \mathbb{R}$ denotes time, and $u:=u(z,t)$.

%

\begin{figure}[h!]
	\centering	
	\includegraphics[scale = 0.6]{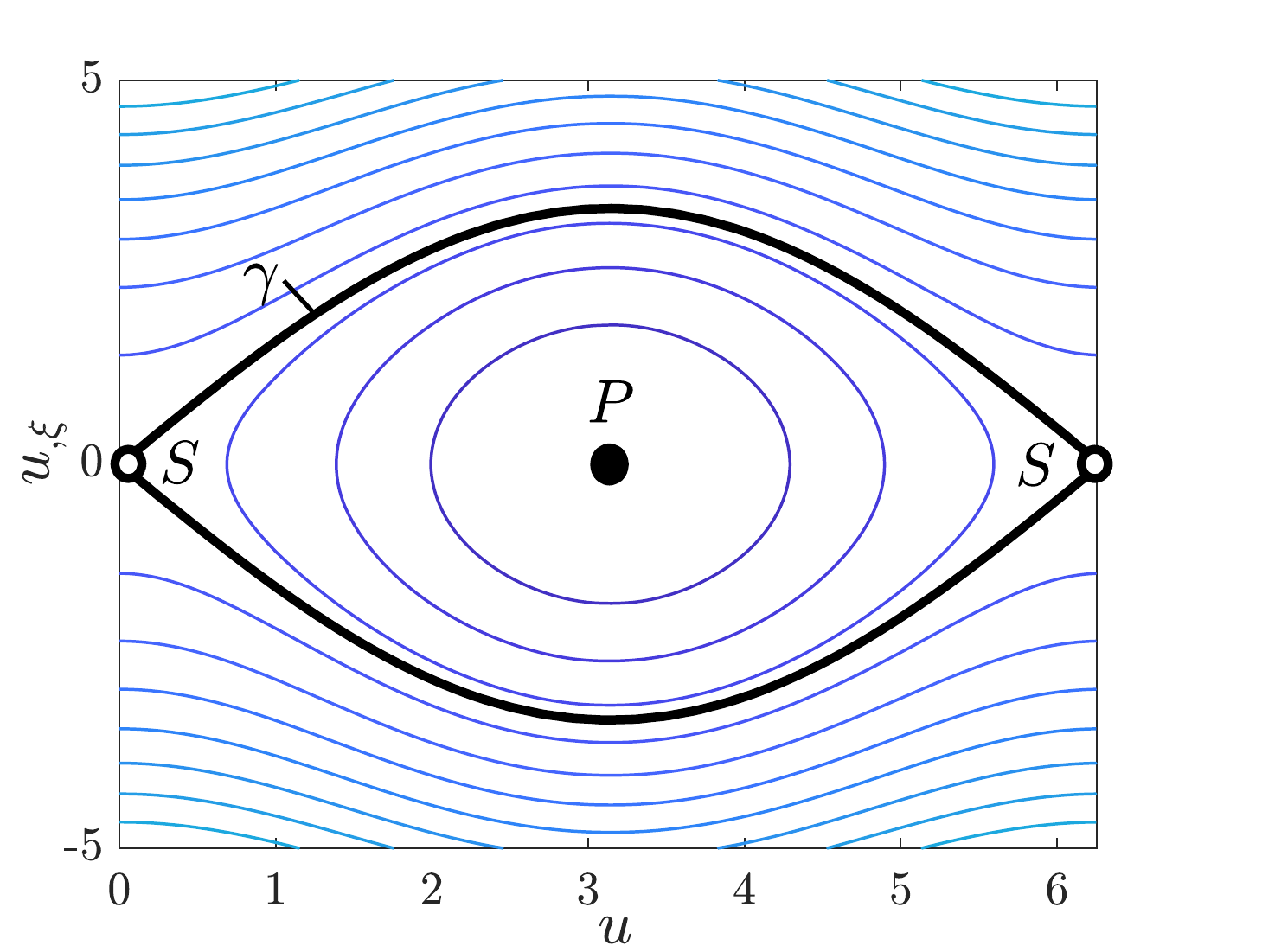}
	\caption{Homoclinic orbit $\gamma$ for the ordinary differential equation~\eqref{eq:sineGordon_stationary} with saddle point $S$ and fixed point $P$.}
	\label{fig:homoclinicOrbitPhase_sine}%
\end{figure}

Using the same procedure as in Sec.~\ref{sec:StationaryWaveSolutions}, we apply the transformation $\xi=z-ct$ in order to determine stationary solutions of the sine-Gordon equation~\eqref{eq:sineGordon}. With this transformation we obtain from Eq.~\eqref{eq:sineGordon} the following ordinary differential equation

\begin{equation}\label{eq:sineGordon_stationary}
u_{,\xi\xi}+\frac{1}{c^2-1} \sin u=0.
\end{equation}
For Eq.~\eqref{eq:sineGordon_stationary}, the homoclinic orbit on a cylindrical phase space, which is $2\pi$-periodic with respect to $u$, is given by~(cf., \cite{dostal:2017bneu})
\begin{equation}\label{eq:sineGordon_gamma}
\begin{aligned}
    \gamma(u,u_{,\xi})= \left\{u\in [0,2\pi), u_{,\xi} \in \mathbb{R},|c|<1: u_{,\xi}^2− \frac{2}{c^2-1} \cos u = -\frac{2}{c^2-1}\right\}.\\
\end{aligned}
\end{equation}
This homoclinic orbit is shown in Fig.~\ref{fig:homoclinicOrbitPhase_sine}.


Starting at the upper or lower part of this homoclinic orbit close to the saddle point $S$, and solving Eq.~\eqref{eq:sineGordon_stationary} numerically, we obtain the so-called kink or anti-kink solution, respectively. 
By solving Eq.~\eqref{eq:sineGordon_stationary} one can also show that the analytical expression for the kink and anti-kink solutions is given by
\begin{equation}\label{eq:sineGordon_kink_analytical}
u_{\mathrm{k}}(\xi)=4\arctan\left( \pm\exp\left(\frac{\xi-\xi_0}{\sqrt{1-c^2}}\right)\right),
\end{equation}
where $\xi_0$ denotes the centre position.

\begin{figure}[h!]
	\centering	
	\includegraphics[scale = 0.6]{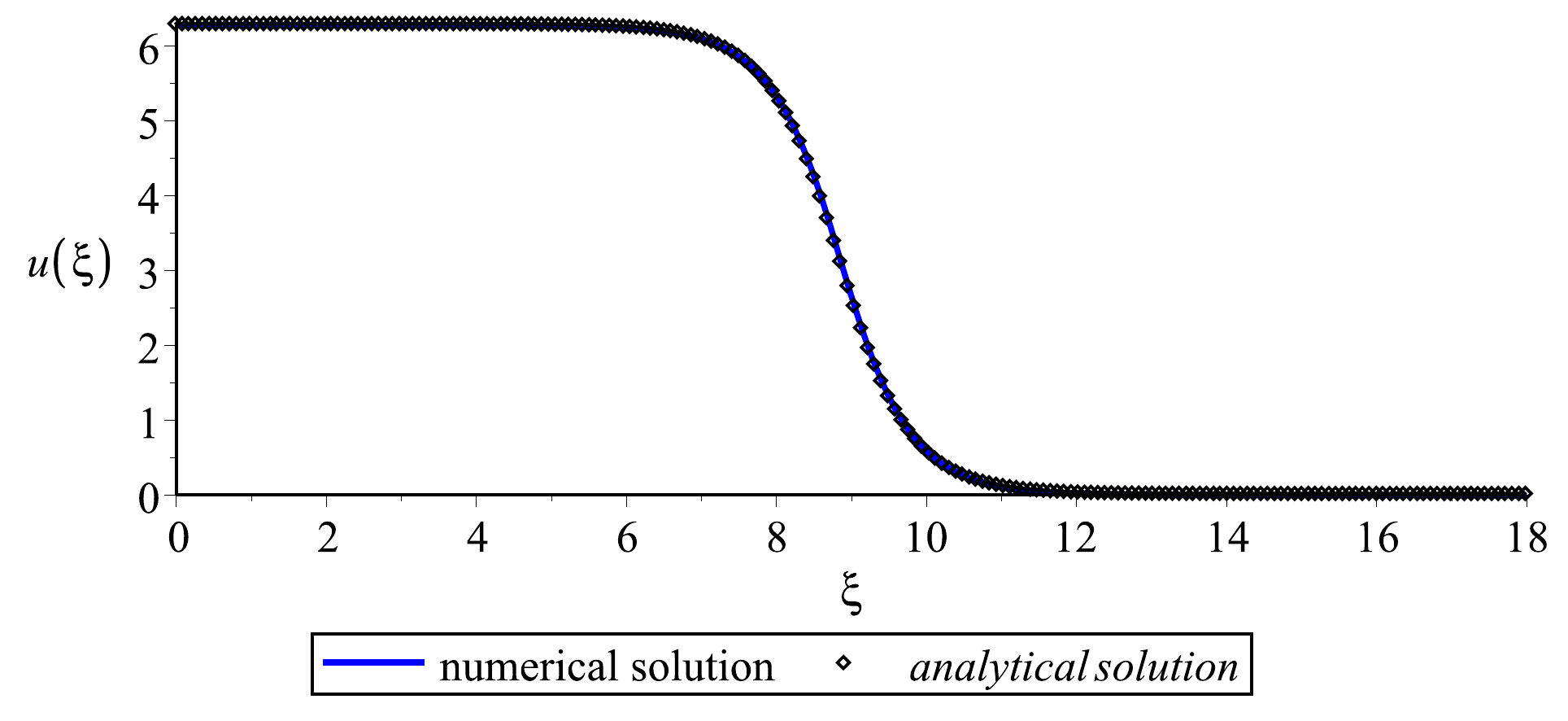}
	\caption{Anti-kink solution for the sine-Gordon equation, where $c=0.8$.}
	\label{fig:kink_solution_sine}%
\end{figure}

A numerically obtained anti-kink solution of the sine-Gordon~Eq.~\eqref{eq:sineGordon_stationary} is shown in Fig.~\ref{fig:kink_solution_sine} for the case $c=0.8$, and it clearly coincides with the corresponding analytical solution from Eq.~\eqref{eq:sineGordon_kink_analytical}.
 These kink or anti-kink solutions are localized stationary solutions of the sine-Gordon Eq.~\eqref{eq:sineGordon}. Similarly as for the equation of motion (Eq.~\eqref{eq:nonlinEoM2}) of the nonlinear gradient elasticity model, they are located at a homoclinic orbit of the corresponding ordinary differential equation for the transformed coordinate $\xi=z-ct$.

\bibliographystyle{model1-num-names}
\bibliography{book_lit,paper_lit,library}








\end{document}